\documentclass[twocolumn,twocolappendix,usenames,dvipsnames]{aastex63}

\usepackage{amsmath}

\shorttitle{Spatial Decorrelation of Young Stars and Dense Gas}
\shortauthors{Semenov, Kravtsov, Gnedin}
\graphicspath{{./}{figures/}}

\usepackage{etoolbox}
\makeatletter
\patchcmd{\NAT@citex}
  {\@citea\NAT@hyper@{\NAT@nmfmt{\NAT@nm}\NAT@date}}
  {\@citea\NAT@nmfmt{\NAT@nm}\NAT@hyper@{\NAT@date}}
  {}
  {}
\patchcmd{\NAT@citex}
  {\@citea\NAT@hyper@{%
     \NAT@nmfmt{\NAT@nm}%
     \hyper@natlinkbreak{\NAT@aysep\NAT@spacechar}{\@citeb\@extra@b@citeb}%
     \NAT@date}}
  {\@citea\NAT@nmfmt{\NAT@nm}%
   \NAT@aysep\NAT@spacechar%
   \NAT@hyper@{\NAT@date}}
  {}
  {}
\patchcmd{\NAT@citex}
  {\@citea\NAT@hyper@{%
     \NAT@nmfmt{\NAT@nm}%
     \hyper@natlinkbreak{\NAT@spacechar\NAT@@open\if*#1*\else#1\NAT@spacechar\fi}%
       {\@citeb\@extra@b@citeb}%
     \NAT@date}}
  {\@citea\NAT@nmfmt{\NAT@nm}%
   \NAT@spacechar\NAT@@open\if*#1*\else#1\NAT@spacechar\fi%
   \NAT@hyper@{\NAT@date}}
  {}
  {}
\makeatother

\begin{document}

\title{Spatial Decorrelation of Young Stars and Dense Gas\\ as a Probe of the Star Formation--Feedback Cycle in Galaxies}

\author[0000-0002-6648-7136]{Vadim A. Semenov}
\altaffiliation{\href{mailto:vadim.semenov@cfa.harvard.edu}{vadim.semenov@cfa.harvard.edu} \\ NHFP Hubble Fellow.}
\affiliation{Center for Astrophysics $|$ Harvard \& Smithsonian, 60 Garden St, Cambridge, MA 02138, USA}

\author[0000-0003-4307-634X]{Andrey V. Kravtsov}
\affiliation{Department of Astronomy \& Astrophysics, The University of Chicago, Chicago, IL 60637, USA}
\affiliation{Kavli Institute for Cosmological Physics, The University of Chicago, Chicago, IL 60637, USA}
\affiliation{Enrico Fermi Institute, The University of Chicago, Chicago, IL 60637, USA}

\author[0000-0001-5925-4580]{Nickolay Y. Gnedin}
\affiliation{Department of Astronomy \& Astrophysics, The University of Chicago, Chicago, IL 60637, USA}
\affiliation{Kavli Institute for Cosmological Physics, The University of Chicago, Chicago, IL 60637, USA}
\affiliation{Fermilab Center for Particle Astrophysics, Fermi National Accelerator Laboratory, Batavia, IL 60510-0500, USA}

\def\Mnsf{{M}_{\rm nsf}}
\def\Msf{{M}_{\rm sf}}
\def\Mg{{M}_{\rm g}}
\def\Mst{{M}_\star}
\def\MH2{M_{\rm H_2}}

\def\SFR{\dot{M}_\star}
\def\dotMsf{\dot{M}_{\rm sf}}
\def\Fsf{F_{\rm sf}}
\def\Fp{F_{\rm +}}
\def\Fm{F_{\rm -}}
\def\Fmdyn{F_{\rm -,d}}
\def\Fmfb{F_{\rm -,fb}}

\def\rhoSFR{{\dot{\rho}}_\star}
\def\rhoH2{\rho_{\rm H_2}}
\def\epsff{\epsilon_{\rm ff}}
\def\epsffcr{\varepsilon_{\rm ff,cr}}
\def\epsint{\epsilon_{\rm tot}}
\def\avir{\alpha_{\rm vir}}
\def\avirsf{\alpha_{\rm vir,sf}}
\def\cs{c_{\rm s}}
\def\st{\sigma_{\rm t}}
\def\stot{\sigma_{\rm tot}}
\def\rhosf{\rho_{\rm sf}}
\def\nsf{n_{\rm sf}}
\def\Tsf{T_{\rm sf}}

\def\Sg{\Sigma_{\rm g}}
\def\Sstar{\Sigma_{\rm \star}}
\def\Smolres{\Sigma_{\rm H_2,40\pc}}
\def\Smolkpc{\Sigma_{\rm H_2,1\kpc}}
\def\Ssf{\Sigma_{\rm sf}}
\def\SSFRres{\dot\Sigma_{\star{\rm ,40\pc}}}
\def\SSFRkpc{\dot\Sigma_{\star{\rm ,1\kpc}}}

\def\fsf{f_{\rm sf,H_2}}
\def\fH2{f_{\rm H_2}}
\def\fgas{f_{\rm g}}

\def\b{b}
\def\xieff{\xi_{\rm eff}}

\def\Nc{N_{\rm c}}
\def\tsf{t_{\rm sf}}
\def\tnsf{t_{\rm nsf}}
\def\tc{t_{\rm c}}
\def\tH2{\tau_{\rm H_2}}
\def\tL{\tH2(L)}
\def\tdep{t_\star}
\def\taust{\tau_\star}
\def\tp{\tau_{+}}
\def\tm{\tau_{-}}
\def\tmd{\tau_{\rm -,d}}
\def\tmfb{\tau_{\rm -,fb}}
\def\tff{t_{\rm ff}}
\def\tauff{\tau_{\rm ff}}
\def\taufflow{\tau_{\rm ff}^{\rm dr}}
\def\tauffhi{\tau_{\rm ff}^{\rm sr}}
\def\tsfr{\Delta t}
\def\tsfe{\tau_{\rm e,sf}}
\def\tmH2{t_{\rm H_2}}
\def\tMC{t_{\rm MC}}
\def\tmolnsf{t_{\rm H_2,nsf}}

\def\sigtot{\sigma_{\rm tot}}
\def\vc{v_{\rm c}}

\def\pc{{\rm\;pc}}
\def\kpc{{\rm\;kpc}}
\def\Myr{{\rm\;Myr}}
\def\Gyr{{\rm\;Gyr}}
\def\Msun{{\rm\;M_\odot}}
\def\Msunyr{{\rm\;M_\odot\;yr^{-1}}}
\def\Msunpc2{{\rm\;M_\odot\;pc^{-2}}}
\def\kms{{\rm\;km\;s^{-1}}}
\def\cc{{\rm\;cm^{-3}}}
\def\mpcc{\;m_{\rm p}{\rm\;cm^{-3}}}
\def\K{{\rm\;K}}

\def\ns{$n$--$\sigtot$ }
\def\const{{\rm const}}
\def\Lstar{$L_\star$}

\def\Fmolp{\Fp^{\rm H_2}}
\def\Fmolm{\Fm^{\rm H_2}}
\def\tmolp{\tp^{\rm H_2}}
\def\tmolm{\tm^{\rm H_2}}

\def\q{\mathbf{q}}

\def\UMW{U_{\rm MW}}

\def\taudep{\tau_{\rm dep}}
\def\tauH2{\tau_{\rm dep,H_2}}
\def\tauHI{\tau_{\rm dep,HI}}

\def\SSFR{\dot\Sigma_\star}
\def\SH2{\Sigma_{\rm H_2}}
\def\SHI{\Sigma_{\rm HI}}

\def\eturb{e_{\rm turb}}

\def\Halpha{H$\alpha$}

\begin{abstract}
The spatial decorrelation of dense molecular gas and young stars observed on $\lesssim 1$ kiloparsec scales in nearby galaxies indicates rapid dispersal of star-forming regions by stellar feedback. We explore the sensitivity of this decorrelation to different processes controlling the structure of the interstellar medium, the abundance of molecular gas, star formation, and feedback in a suite of simulations of an isolated dwarf galaxy with structural properties similar to NGC\,300 that self-consistently model radiative transfer and molecular chemistry. Our fiducial simulation reproduces the magnitude of decorrelation and its scale dependence measured in NGC\,300, and we show that this agreement is due to different aspects of feedback, including H$_2$ dissociation, gas heating by the locally variable UV field, early mechanical feedback, and supernovae. In particular, early radiative and mechanical feedback affects the correlation on $\lesssim 100$ pc scales, while supernovae play a significant role on $\gtrsim 100$ pc scales. The correlation is also sensitive to the choice of the local star formation efficiency per freefall time, $\epsilon_{\rm ff}$, which provides a strong observational constraint on $\epsilon_{\rm ff}$ when the global star formation rate is independent of its value. Finally, we explicitly show that the degree of correlation between the peaks of molecular gas and star formation density is directly related to the distribution of the lifetimes of star-forming regions.  
\end{abstract}


\section{Introduction} \label{sec:intro}

Studies over the past decade have clearly demonstrated that modeling of the star formation--feedback cycle is a key ingredient that shapes the properties of galaxies in cosmological simulations \citep[e.g.,][see also \citealt{Naab.Ostriker.2017} and \citealt{Vogelsberger.etal.2020} for recent reviews]{Governato.etal.2010,Brook.etal.2012,Agertz.etal.2013,Hopkins.etal.2013,Hopkins.etal.2014,Stinson.etal.2013,Agertz.Kravtsov.2015,Agertz.Kravtsov.2016}. 

Although the sophistication of models for star formation, stellar feedback, and the thermodynamics and chemistry of the interstellar medium (ISM) has experienced dramatic progress in the past several years \citep[e.g.,][]{Robertson.Kravtsov.2008,Gnedin.Kravtsov.2010,Gnedin.Kravtsov.2011,Hopkins.etal.2011,Hopkins.etal.2012,Christensen.etal.2012,Kannan.etal.2014,Kannan.etal.2020,Marinacci.etal.2019,Benincasa.etal.2020,Smith.etal.2020}, there are still significant theoretical uncertainties in the relevant physical processes and their specific numerical implementation in subgrid recipes. Given that a number of key properties of galaxies can be sensitive to these processes, this  poses substantial challenges for galaxy formation modeling \citep[][]{Keller.etal.2019,Munshi.etal.2019}. Thus, the calibration of such models using observations is often required. To validate such models, their results should be confronted with observations not used in model calibrations \citep[e.g.,][]{Grisdale.etal.2017}.  

New high-resolution observations of star-forming regions and dense molecular gas serve as an important testing ground for a new generation of sophisticated high-resolution models \citep[e.g.,][]{Benincasa.etal.2013,Buck.etal.2019,Li.etal.2020,Grisdale.2021}. Indeed, as galaxy simulations reach spatial resolutions of $\lesssim 10-100$ pc, a number of new observational surveys have been conducted to probe the distribution of young stars and dense molecular gas on comparable scales \citep[e.g.,][]{Meidt.etal.2013,Meidt.etal.2020,Faesi.etal.2014,Faesi.etal.2016,Faesi.etal.2018,Leroy.etal.2016, Leroy.etal.2017,Schruba.etal.2017,Sun.etal.2018,Sun.etal.2020,Querejeta.etal.2019,Schinnerer.etal.2019,Lee.etal.2021}.

An example of a new generation of observational probes into the star formation--feedback cycle is the spatial decorrelation between peaks in the spatial  distribution of young massive stars traced via \Halpha~emission and peaks of molecular gas traced by its CO emission. The distributions of \Halpha~and CO can now  be mapped with sub-100 pc resolution in a sample of nearby galaxies 
\citep[e.g.,][]{Kreckel.etal.2018,Kruijssen.etal.2019,Schinnerer.etal.2019,Chevance.etal.2020,Chevance.etal.2021}.
This decorrelation can be quantified by measuring the depletion time of molecular gas---defined as the ratio of molecular mass and star formation rate (SFR) within a patch, $\tauH2=\MH2/\SFR$---in patches of different size and centered on peaks of either CO or \Halpha~emission.

As first shown by \citet{Schruba.etal.2010} for the M33 galaxy, when patch sizes are smaller than a kiloparsec, the depletion time of gas in the patches centered on \Halpha~peaks is several times \emph{shorter} than the global depletion time in the galaxy because such patches preferentially include the tracer of recent star formation. Conversely, in the patches of the same size centered on the CO peaks, $\tauH2$ is several times \emph{longer} than the global value. As the patch size is increased, the differences diminish until the depletion times in both types of centering converge to the global value for patch sizes $\gtrsim 1$ kpc. The characteristic shape of the divergence of $\tauH2$ in the patches centered on CO- and \Halpha~peaks with decreasing patch size is reminiscent of a tuning fork, and the corresponding plot has been dubbed \emph{``the tuning fork diagram''} \citep{Kruijssen.etal.2018,Chevance.etal.2020}, which we will also use as a shorthand term in this paper.\footnote{Note that the tuning fork diagram explored here is distinct from the ``Hubble Tuning Fork,'' which is often used to classify galaxy morphologies.} 

In general, star formation is expected to occur in the cold, dense molecular gas \citep[e.g.,][]{Kennicutt.Evans.2012}. Therefore, the decorrelation between dense gas and young stars is most likely a signature of rapidly operating feedback processes in and around star-forming regions. Indeed, the existence of isolated \Halpha~peaks already presumes a feedback process that ionizes the gas on a short timescale. The overall dependence of depletion time on scale, however, likely bears an imprint of all the collective feedback processes that operate in the region, including the effects of spatial correlations of star formation sites.

An early attempt to model such decorrelation of molecular gas and young stars on small scales in the context of scatter of the molecular depletion time as a function of scale was done by \citet{Feldmann.etal.2012}, who showed that such measurements can be used as a probe of stochasticity of star formation in individual regions. More recently, 
\citet{Fujimoto.etal.2019} used simulations of an isolated Milky Way-size galaxy and compared the estimates of $\tauH2$ in the \Halpha- and CO-centered peaks as a function of scale to the measurements in NGC\,300 \citep[][]{Kruijssen.etal.2019}. Although their simulations included most of the processes thought to be critical for star formation and feedback modeling, these authors found that in their simulation almost all young stellar emission was
associated with molecular CO emission at all scales down to $\approx 50$ pc. They attributed this failure to match the strong observed trend to inadequate modeling of presupernova feedback, in particular to insufficient realism of effects of photoionization feedback. 

This conclusion is consistent with the interpretation of the observed decorrelation of CO- and \Halpha-emitting gas in nine nearby galaxies by 
\citet{Chevance.etal.2020,Chevance.etal.2021}, who concluded that their observations indicate that molecular gas is dissociated and/or  dispersed on average within 3 Myr after a peak in young stars becomes visible at the optical wavelength (i.e., roughly $\sim 5\text{--}6$ Myr since the onset of local star formation). 
Observational measurements of the spatial decorrelation as a function of patch size can thus be used as a probe of early stellar feedback and as a test of its modeling in galaxy formation simulations.

Conversely, simulations that reproduce the scale dependence of depletion times can also provide insights for interpretation of observations. For example, the short evolution timescales of star-forming regions derived by \citet{Kruijssen.etal.2019} and \citet{Chevance.etal.2020,Chevance.etal.2021} agree \emph{quantitatively} with the predictions of hydrodynamic simulations of galaxies, where these timescales can be measured directly by following the evolution of ISM gas parcels between different states \citep{Semenov.etal.2017,Semenov.etal.2019} or tracking giant molecular clouds \citep[e.g.,][]{Grisdale.etal.2019,Benincasa.etal.2020b,Jeffreson.etal.2021}. The simulations, however, show that these timescales are not the same for all star-forming regions but exhibit a broad distribution. Such simulations then can be used to elucidate the connection between the details of these distributions and the scale dependence of depletion times.

With these motivations in mind, we explore the scale dependence of depletion times in galaxy simulations with
a successful implementation of a star formation and feedback model that we recently used to understand the origin of long depletion times in galaxies \citep[][]{Semenov.etal.2017, Semenov.etal.2018} and approximate linearity of the molecular Kennicutt--Schmidt relation \citep[][]{Semenov.etal.2019}. \citet{Semenov.etal.2018} showed (see their Figure~11) that 
simulations of a Milky Way-size galaxy can reproduce the scale dependence of $\tauH2$ at scales $L\gtrsim 75$ pc measured by \citet{Schruba.etal.2010}. 

Here we present a detailed analysis of this dependence in a suite of simulations of a galaxy with structural properties closely matching those of NGC\,300. NGC\,300 is a nearby \citep[$\sim 2$ Mpc; e.g.,][]{Gieren.etal.2005,Rizzi.etal.2006} sub-\Lstar~galaxy seen at a favorable inclination angle. This galaxy is massive enough to sustain a thin gaseous disk, while at the same time, the effects of stellar feedback are more pronounced than in more massive Milky Way-like galaxies. All these factors make NGC\,300 an ideal laboratory for observational studies of star formation and feedback \citep[e.g.,][]{Deharveng.etal.1988,Faesi.etal.2014,Faesi.etal.2016,Faesi.etal.2018,McLeod.etal.2020} and, in particular, the scale dependence of depletion times \citep{Kruijssen.etal.2019}.

We focus on exploring the sensitivity of $\tauH2$ scale dependence to star formation and feedback modeling. To this end, we vary the assumptions and parameters of these models; in particular, we explicitly test the effects of self-consistent modeling of radiative transfer (RT) and photoionization of the natal star-forming region by young massive stars.
By closely matching the structural properties of NGC\,300 we avoid any global effect of, e.g., galaxy mass, size, gas fraction, and so forth on the scale dependence of $\tauH2$ and can directly compare our results with the observational findings of \citet{Kruijssen.etal.2019}. Although the observed galaxy-to-galaxy variation of this statistic is relatively small, its details do change with global galaxy properties, thereby affecting the conclusions about the relative importance of different ISM and feedback processes \citep{Chevance.etal.2020}.

The paper is organized as follows. In Section~\ref{sec:sims} we describe our simulations and compare the bulk properties and radial profiles of the model galaxy to the observed properties of NGC\,300. We also describe the details of how CO and SFR peaks are identified in our analysis and how molecular depletion time is measured in patches of different scales. We present the results of our fiducial model in Section~\ref{sec:results:tf-fid}, showing that it matches the observed decorrelation as a function of scale quite well, and explore the sensitivity of the results to variations of feedback and star formation modeling in the rest of Section~\ref{sec:results}. We discuss our results in Section~\ref{sec:discussion} and summarize conclusions in Section~\ref{sec:summary}.

\section{Simulations} 
\label{sec:sims}

\subsection{Simulation Code Overview}
\label{sec:sims:code}

To simulate our NGC\,300-like galaxy, we use the adaptive mesh refinement (AMR) $N$-body and gasdynamics code ART \citep{Kravtsov.1999,Kravtsov.etal.2002,Rudd.etal.2008,Gnedin.Kravtsov.2011} with self-consistent modeling of RT \citep[][]{Gnedin.2014}.
The hydrodynamic fluxes in the ART code are handled by a second-order Godunov-type method \citep{Colella.Glaz.1985} with a piecewise linear reconstruction of states at the cell interfaces \citep{vanLeer.1979} and a monotonized central slope limiter based on \citet{Colella.1985}. The Poisson equation for the gravitational potential of gas, stars, and dark matter is solved by using a Fast Fourier Transform at the lowest grid level and relaxation method on all higher refinement levels, with the effective resolution for gravity corresponding to $\sim$2--4 cells \citep[see][]{Kravtsov.etal.1997,Gnedin.2016,Mansfield.Avestruz.2020}. 
The AMR grid is adaptively refined when the gas mass in a cell exceeds $\sim2\,500 \Msun$, reaching the maximal resolution of $\Delta = 10\pc$ that matches the resolution of observations used in our comparison \citep[$\sim 20\pc$;][]{Kruijssen.etal.2019}.

To model the relation between molecular gas and young stars as realistically as possible, we include a number of key processes affecting the formation and destruction of molecular gas as well as a physically motivated model for star formation. The processes modeled in our fiducial simulation are detailed below, together with the parameter variations that we explore.

\emph{Radiative transfer} of UV field is modeled self-consistently using the Optically Thin Variable Eddington Tensor approximation \citep[OTVET;][]{Gnedin.Abel.2001,Gnedin.2014}. The ionizing radiation field is sampled at the ionization thresholds for \ion{H}{1}, \ion{He}{1}, and \ion{He}{2} and includes the contribution from both the local sources and the \citet{Haardt.Madau.2012} cosmological background at redshift $z=0$. To model H$_2$ photodissociation, we also model RT in the Lyman--Werner bands as described in \citet{Ricotti.etal.2002}. To test the effect of the time-dependent and spatially inhomogeneous radiation field, we also rerun our simulation without RT, using a uniform UV background specified below.

\emph{Gas heating and cooling} are treated using the method of \citet{Gnedin.Hollon.2012} with the metallicity-dependent part of the cooling and heating functions dependent on the radiation field that can arbitrarily vary in time and space. The cooling and heating rates in this approximation are parameterized via seven numbers: the gas density, temperature, and metallicity as well as the photoionization rates of \ion{H}{1}, \ion{He}{1}, and \ion{C}{6} and the photodissociation rate of H$_2$ in the Lyman--Werner bands. The latter four rates parameterize local variations of the radiation field at different energies, with $\Gamma_{\rm LW}$, $\Gamma_{\rm HI}$, and $\Gamma_{\rm HeI}$ describing the field at $11.2\text{--}24.6$ eV and $\Gamma_{\rm CVI}$ sampling high-energy photons at $470$ eV.
In our simulations with RT, all these rates are computed self-consistently from the local radiation field. The metallicity-independent part of the cooling and heating functions is computed exactly by summing over all relevant reactions involving H and He ions and molecular hydrogen, without assuming ionization equilibrium \citep[see Appendix A.4 in][]{Gnedin.Kravtsov.2011}. 

In the resimulation without RT, we adopt the constant UV background with the average photoionization rates from the ISM of our RT simulation: ($\Gamma_{\rm LW}$, $\Gamma_{\rm HI}$, $\Gamma_{\rm HeI}$, $\Gamma_{\rm CVI}$) = ($2\times10^{-11}$, $2\times10^{-17}$, $3\times10^{-16}$, $9\times10^{-18}$) ${\rm s^{-1}}$. To account for the shielding of dense gas from the background radiation, we use a prescription calibrated in RT simulations of the ISM \citep[the “L1a” model in][]{Safranek-Shrader.etal.2017}. Interestingly, we find that despite strong attenuation, the photoionization rates have a strong effect on the NGC\,300 outskirts. Resimulation of this galaxy with $\Gamma_{\rm HI}$, $\Gamma_{\rm HeI}$, and $\Gamma_{\rm CVI}$ all set to 0 and $\Gamma_{\rm LW}=2\times10^{-11}\;{\rm s^{-1}}$ leads to an excessive heating at $R>3\kpc$ and a substantially smaller star-forming and molecular disk.

In addition, we find that heating by X-rays from the cosmic background also has a strong effect on the NGC\,300 outskirts. The tables of cooling and heating rates from \citet{Gnedin.Hollon.2012} are not wide enough to properly capture gas cooling and heating in this regime. By running additional Cloudy models we found that we can compensate for this deficiency by ignoring the absorption of X-rays from the cosmic background only (while treating all stellar radiation self-consistently). We use this numerical hack in all RT simulations presented in this paper.

\emph{Molecular chemistry} is computed on the fly by using the ``six-species model’’ described in the appendix of \citet{Gnedin.Kravtsov.2011} that explicitly tracks the evolution of \ion{H}{1}, \ion{H}{2}, \ion{He}{1}, \ion{He}{2}, \ion{He}{3}, and H$_2$ on the AMR grid, coupled with the local radiation field. After a series of experiments, we made two modifications to the H$_2$ modeling: we added a ceiling on the size of the shielded regions (estimated using the Sobolev approximation) of 100 pc and reduced the clumping factor of H$_2$ from 10 to 3. The second change is motivated by the higher resolution of our simulations: since they resolve a larger range of spatial scales than simulations of \citet{Gnedin.Kravtsov.2011}, the contribution to the clustering of H$_2$ gas from the unresolved scales is reduced proportionately.

Self-consistently computed H$_2$ densities are only available in the RT simulation, while the simulations without RT require a model for $\rho_{\rm H_2}$ and the assumption about the incident radiation field. To this end, we use the parameterization from \citet[][Equations~(6)--(7)]{Gnedin.Kravtsov.2011}:

\begin{align}
\rho_{\rm H_2} &= \frac{\rho_{\rm H}}{1+\exp{(-4x-3x^3)}},\\
     x &\equiv \Lambda^{3/7} \log \left( Z \frac{n_{\rm H}}{\Lambda n_\star} \right),
\end{align}
where $\rho_{\rm H} = 0.76\rho$ and $n_{\rm H}=\rho_{\rm H}/m_{\rm p}$ are the volume and number density of all hydrogen (assuming 0.24 mass fraction of helium and all heavier elements), $Z$ is gas metallicity, and $n_\star$ and $\Lambda$ are tunable parameters that encode the dependence on the radiation field and ISM structure. 
These parameters can be calibrated using RT simulations, as was done in \citet{Gnedin.Kravtsov.2011} and \citet{Gnedin.Draine.2014}. However, we find that neither of these calibrations can reproduce the results of our RT simulation of NGC\,300, indicating that the spatial resolution, the star formation and feedback model, and overall structure of the ISM in this galaxy are substantially different from the simulations of the dense $z=3\text{--}4$ gas-rich disk used in prior calibrations. Therefore, we recalibrate these parameters specifically for our simulated galaxy and use the values of $n_\star = 60\cc$ and $\Lambda=0.5$ that reproduce the radial profile of H$_2$ surface density inside $R<3\kpc$---the region where we perform our analysis---with deviations of $<20\%$ from the full H$_2$ chemistry results.

\emph{Supernova (SN) and mechanical pre-SN feedback.} In addition to radiative feedback, young stars in our simulations also inject thermal energy and radial momentum following our fiducial model from \citet{Semenov.etal.2017,Semenov.etal.2018,Semenov.etal.2019}. The amount of energy and radial momentum injected per SN are computed using the fits to simulations of SN remnants evolution in a nonuniform ISM by \citet{Martizzi.etal.2015}. In our fiducial model, we additionally boost the radial momentum by a factor of 5 to account for the effects of SN clustering \citep[e.g.,][]{Gentry.etal.2017,Gentry.etal.2018} and cosmic ray pressure \citep{Diesing.Caprioli.2018}, both of which can increase the injected momentum by a factor of a few. To test the effect of the total feedback momentum budget, we also explore the case without such a boost. 
The total number of SNe for a given star particle is computed using the \citet{Chabrier.2003} IMF.

Young stars can also affect the ISM via stellar winds, pressuring \ion{H}{2} regions, and dust-reprocessed radiation pressure before the first SN explosions---the processes often referred to collectively as ``early feedback.’’ As the momentum injection rate due to early feedback processes is approximately the same as that of the SNe \citep[e.g.,][]{Agertz.etal.2013}, we approximate the effects of early feedback by starting momentum injection from the moment when the stellar particle is formed, without any delay before the first SN explosion, and continue the injection for 40 Myr. 
To test the relative roles of early feedback and SNe, we also resimulated our galaxy with two additional feedback models: (i) without any pre-SN feedback, by introducing the delay before momentum injection of 3 Myr, and (ii) without SNe, by injecting momentum at the same fiducial rate but only during the first 3 Myr.
Note that because of the difference in the injection duration, the total feedback budget in (ii) is reduced by a factor of $3/40$.

\emph{Subgrid turbulence model.} Another important feature of our simulations is the explicit dynamic modeling of unresolved turbulence. Our implementation is based on the ``shear-improved’’ model of \citet{Schmidt.etal.2014} and detailed in \citet{Semenov.etal.2016}. In this model, the unresolved turbulent energy, $\eturb$, is sourced by the fluctuating part of the resolved velocity field and decays on the timescale close to the turbulence turnover time on the scale of the cell size.
Advection and the $PdV$ work done by turbulence are treated using the entropy-conserving scheme described in Appendix A of \citet{Semenov.etal.2020}.
Unresolved turbulence provides a nonthermal pressure support and, most importantly, directly couples with the star formation prescription as described below. 

\emph{Star formation prescription.} We use the common parameterization for the local SFR via the star formation efficiency per freefall time, $\tff = \sqrt{3\pi/32G\rho}$:
\begin{equation}
\label{eq:rhosfr}
\rhoSFR = \epsff \frac{\rho}{\tff}.
\end{equation} 
In our fiducial simulation, we do not adopt any star formation threshold and instead allow $\epsff$ to vary continuously with the local value of the (subgrid) virial parameter following the fit to magnetohydrodynamic simulations of turbulent star-forming regions by \citet{Padoan.etal.2012}:
\begin{equation}
\label{eq:epsff-P12}
\epsff = 0.9 \exp{(-\sqrt{\avir/0.53})},
\end{equation}
with the choice of the prefactor explained in \citet{Semenov.etal.2016}. The virial parameter for each simulation cell with size $\Delta$ is defined as for a uniform sphere with radius $R = \Delta/2$ \citep{Bertoldi.McKee.1992}:
\begin{equation} 
\label{eq:avir}
    \avir \equiv \frac{5 \stot^2 R}{3GM} \approx 13.5 \frac{ (\stot/3\kms)^2 }{ (n/100\cc) (\Delta/10 \pc)^2},
\end{equation}
where $\stot = \sqrt{\st^2+\cs^2}$ accounts for both the unresolved turbulent velocity dispersion, $\st = \sqrt{2\eturb/\rho}$, and thermal support, and the values on the right-hand side reflect the typical conditions in star-forming regions in our NGC\,300 simulations. 

To explore the effect of the star formation prescription on the correlation between young stars and dense gas, we also rerun our RT simulation with a star formation threshold of $\avirsf=10$ and assuming different constant values of $\epsff=1\%$, $3.3\%$, and $10\%$ in gas with $\avir < \avirsf$. We also explored the effect of the star formation threshold choice by using a threshold in gas density of $\nsf=100\cc$ instead of the threshold in $\avir$. This density threshold results in a similar mass fraction of star-forming gas to the simulation with the $\avirsf=10$ threshold. 

Overall, we will present nine simulations with variations of star formation and feedback physics that are summarized in Table~\ref{tab:sims}.  

\def\NA{---}
\newcommand{\setcolwidth}[1]{\multicolumn{1}{>{\centering}p{0.7in}}{#1}}
\begin{deluxetable*}{lcccccc}
\tablecolumns{7}
\tablecaption{Summary of the simulation parameters \label{tab:sims}}
\tablewidth{0pt}
\tablehead{
Label & 
\setcolwidth{Radiative transfer} & 
\setcolwidth{Early mechanical feedback} & 
\setcolwidth{Type II SN feedback} &
\setcolwidth{Momentum injection rate\tablenotemark{a}} & 
\setcolwidth{Star formation threshold} & 
\setcolwidth{Star formation efficiency, $\epsff$}
}
\startdata
\sidehead{\it Fiducial simulation:}
RT & Y & Y & Y & fiducial & \NA & variable $\epsff(\avir)$\tablenotemark{b} \\
\sidehead{\it Variation of the feedback model:}
noRT & N & Y & Y & fiducial & \NA & variable $\epsff(\avir)$ \\
noRT, no early FB & N & N & Y & fiducial & \NA & variable $\epsff(\avir)$ \\
noRT, no SNe & N & Y & N & fiducial & \NA & variable $\epsff(\avir)$ \\
noRT, weak FB & N & Y & Y & reduced & \NA & variable $\epsff(\avir)$ \\
\sidehead{\it Variation of the star formation prescription:}
$\alpha10$; $\epsff=1\%$ & Y & Y & Y & fiducial & $\avir<10$ & $1\%$ \\
$\alpha10$; $\epsff=3.3\%$ & Y & Y & Y & fiducial & $\avir<10$ & $3.3\%$ \\
$\alpha10$; $\epsff=10\%$  & Y & Y & Y & fiducial & $\avir<10$ & $10\%$ \\
$n100$; $\epsff=1\%$  & Y & Y & Y & fiducial & $n>100\cc$ & $1\%$ \\
\enddata
\tablenotetext{a}{The fiducial momentum injection per SN corresponds to the \citet{Martizzi.etal.2015} value boosted by a factor of 5 as described in the text, while the model with reduced momentum does not adopt such a boost.}
\tablenotetext{b}{The value of $\epsff$ continuously varies with local virial parameter in each cell according to Equation (\ref{eq:epsff-P12}).}
\end{deluxetable*}

\subsection{NGC\,300 Galaxy Model}
\label{sec:sims:ics}

\begin{deluxetable}{lcc}
\tablecolumns{3}
\tablecaption{The parameters of the NGC\,300 galaxy used to initialize simulations \label{tab:ics}}
\tablewidth{0pt}
\tablehead{
Parameter & 
Value\tablenotemark{a} & 
Units
}
\startdata
\sidehead{\it NFW dark matter halo:}
Mass, $M_{\rm 200c}$ & $8.3 \times 10^{10}$ & $\Msun$ \\
Concentration, $c_{\rm 200c}$ & 15.4 & \NA \\
\sidehead{\it Exponential stellar disk:}
Mass & $10^{9}$ & $\Msun$ \\
Scale-radius & $1.39$ & $\kpc$ \\
Scale-height & $0.28$ & $\kpc$ \\
\sidehead{\it Exponential gaseous disk:\tablenotemark{b}}
Mass & $2.2 \times 10^{9}$ & $\Msun$ \\
Scale-radius & $3.44$ & $\kpc$ \\
Metallicity at $R = 0$ & 0.76 & $Z_\odot$ \\
Metallicity gradient, $d \log Z / dR$ & $-0.077$ & dex$\;\kpc^{-1}$ \\
\enddata
\tablenotetext{a}{The structural parameters and metallicity gradient are based on the measurements of \citet{Westmeier.etal.2011} and \citet{Bresolin.etal.2009} as described in the text.}
\tablenotetext{b}{The initial scale height of the gaseous disk varies due to disk flaring as it is computed by the \textsc{GalactICS} code assuming constant gas temperature ($\sim 10^4\K$). In the simulation, the disk scale height self-consistently adjusts to the effective pressure gradient of the multiphase ISM, and, therefore, it is not provided in the table.}
\end{deluxetable}

To explore the local effects of star formation and feedback on the small-scale decorrelation of young stars and dense gas, we use an isolated galaxy setup. Such an idealized setup is suitable because the typical timescales of the involved processes ($\lesssim$10 Myr) are significantly shorter than the timescales of any cosmological processes, like accretion of pristine gas from the intergalactic medium and interaction with the circumgalactic medium and nearby galaxies, e.g., other members of the Sculptor Group that NGC\,300 is a part of. In addition, we consider galaxy evolution on the timescales of a few hundred Myr, shorter than the global gas depletion time of a few Gyr. Thus, the gas is not fully exhausted during the simulated time frame. Finally, it is advantageous to study specific processes and their effects in a controlled but realistic setting.

To generate the initial conditions for our NGC\,300 simulations we use the \textsc{GalactICS} code \citep{galactics}. Our simulated galaxy consists of the dark matter halo (modeled with collisionless particles) and exponential stellar and gaseous disks, with the structural parameters of all three components taken from \citet[][W11 hereafter]{Westmeier.etal.2011} and summarized in Table~\ref{tab:ics}. The halo has a Navarro--Frenk--White (NFW) profile with the total mass inside the sphere enclosing an average density of $200 \rho_{\rm cr}$ of $M_{200c}\approx8.3\times10^{10}\Msun$ and the concentration of $c_{\rm 200c}\approx15.4$ (``NFW (fixed)’’ model from Table~3 in W11, which provides a good fit to the observed rotation curve of NGC\,300). The stellar disk has an exponential profile with a scale radius and height of 1.39 and 0.28 kpc, respectively, and a total mass of $10^9 \Msun$. 

The gaseous disk is initialized with an exponential profile with a scale radius of 3.44 kpc and a total mass of $2.2\times10^9\Msun$. Our total gas mass is $\sim10\%$ higher than in W11 because we adjusted it to match the exponential part of the $\SHI$ profile at $R = 5\text{--}10\kpc$ shown in Figure~10 of W11. The observed profile flattens at $R<5\kpc$, which can be due in part to the formation of optically thick and molecular cold neutral medium. On the other hand, a somewhat steeper gas profile adopted in our simulations can be a reason for the mild excess of the atomic and molecular gas and SFR surface densities in the central part of our NGC\,300 analog (see Figures~\ref{fig:prof-vcZ} and \ref{fig:prof-H2SFR} below).
The metallicity of the gaseous disk is initialized using the radial gradient from \citet{Bresolin.etal.2009}: $Z(R) = (0.76\; Z_\odot)\times10^{-0.077R}$, with $R$ in kiloparsecs. 

After we start our simulation, the galaxy undergoes the initial relaxation stage. To mitigate the effect of this initial transient on our results, we first run the galaxy in an adiabatic regime, then turn on cooling, star formation, and mechanical feedback at $t\approx200\Myr$ and, finally, RT and chemistry at $t\approx400\Myr$, so that the changes of physics are more gradual and the galaxy has time to settle down. By $t\approx600\Myr$, all the physics in our fiducial simulation is on and the galaxy is in a quasi-equilibrium state. We then continue our fiducial simulation until 1000 Myr with outputs every 10 Myr, which we use to measure the snapshot-to-snapshot variation of our results. The rest of the runs from Table~\ref{tab:sims} are started from the output of our fiducial simulation at $t\approx600\Myr$ and run until $800\Myr$. The changes in the star formation and feedback model lead to another brief relaxation stage that settles down within $\sim 50\Myr$, and therefore for our analysis, we only use the snapshots at $t = 650\text{--}800\Myr$ from these runs.

\begin{figure*}
\includegraphics[width=\textwidth]{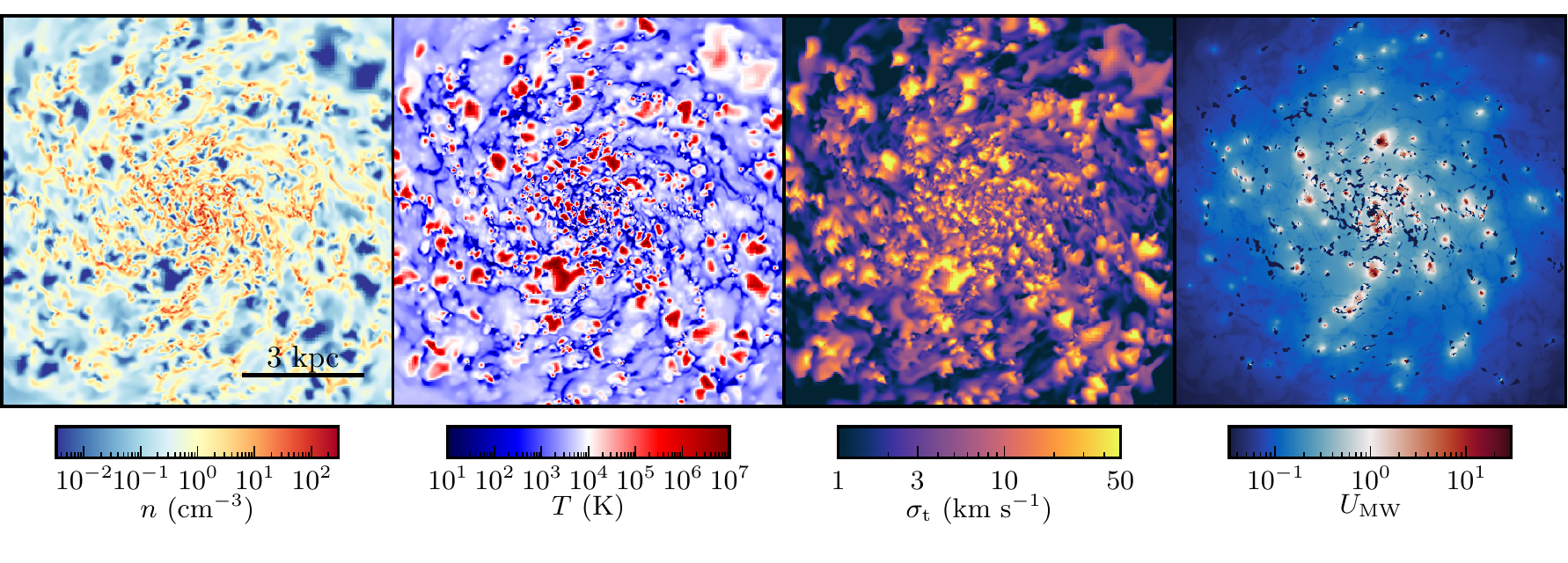}
\caption{\label{fig:maps} Midplane slices of density, temperature, subgrid turbulent velocity, and UV field strength (in Draine units) in our fiducial simulation with RT modeling after 900 Myr of evolution. All these quantities vary by orders of magnitude, and the overall gas structure is highly flocculent, similar to the observed morphology of the ISM in NGC\,300. The ISM structure remains qualitatively similar over the time frame during which we apply our analysis. To illustrate this, an animated version of this figure is available in the online version of the paper. The animation shows the galaxy evolution from 700 to 873 Myr in $\sim$0.25 Myr steps with the real-time duration of 20 seconds.}
\end{figure*}

\begin{figure}
\includegraphics[width=\columnwidth]{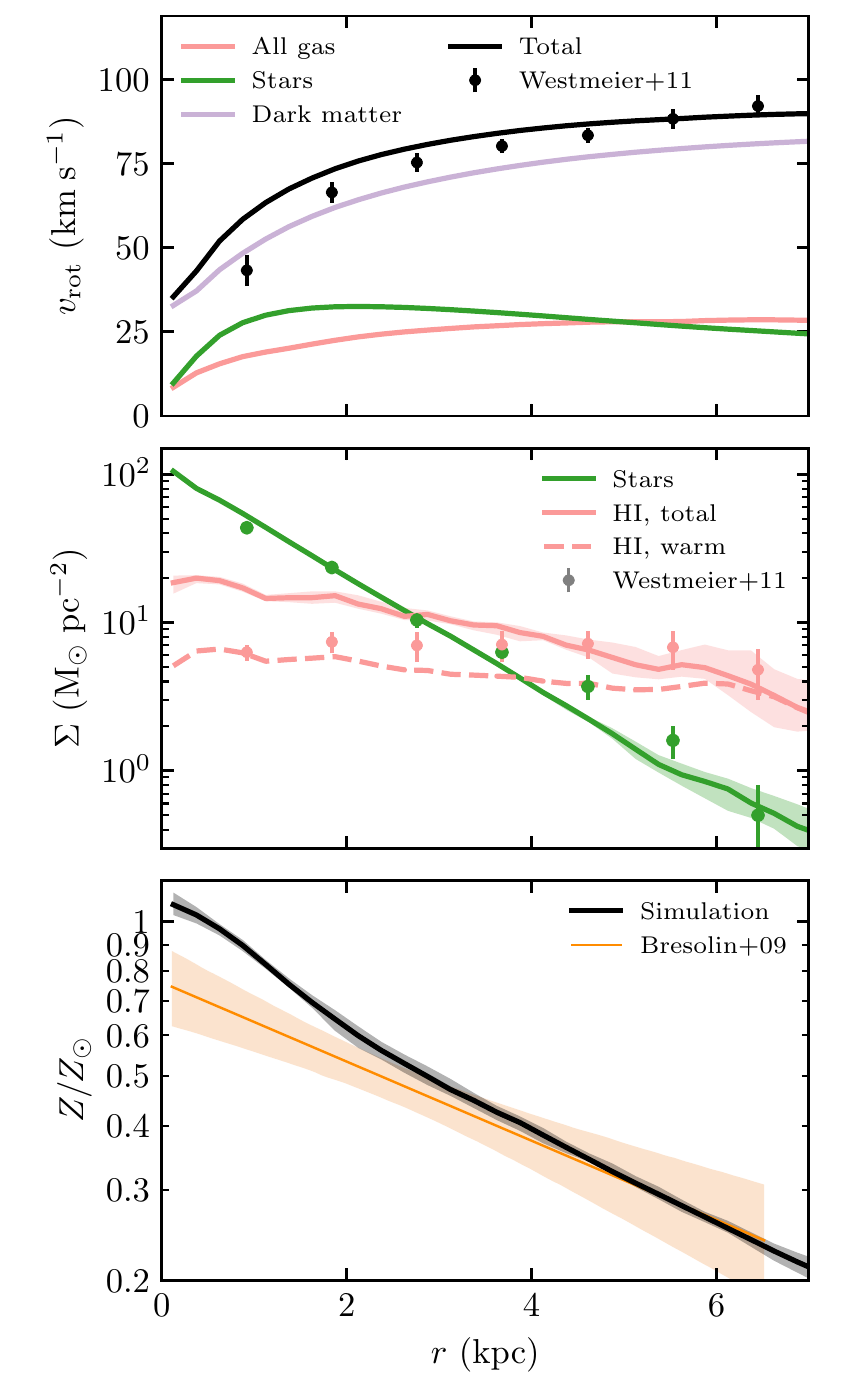}
\caption{\label{fig:prof-vcZ} Comparison of the global profiles from our fiducial simulation with NGC\,300 observations. \emph{Top panel:} profiles of rotational velocity, $v_{\rm rot}$. Simulation lines also show contributions to $v_{\rm rot}$ from gas, stellar disk, and dark matter halo. \emph{Middle panel:} surface density profiles of stellar and \ion{H}{1} disks. To illustrate the possible contribution of the cold optically thick \ion{H}{1}, solid and dashed red lines show the profiles for all atomic gas and only the gas warmer than 1000 K, respectively. \emph{Bottom panel:} gas metallicity profile. 
Simulated profiles are stacked over 41 snapshots between 600 and 1000 Myr, with solid lines showing medians and the shaded region showing the snapshot-to-snapshot variation (16$^{\rm th}$--84$^{\rm th}$ interpercentile range). Points with error bars in the top two panels show the measurements from W11, while the orange band in the bottom panel shows the metallicity gradient measured by \citet{Bresolin.etal.2009}. Our fiducial simulation reproduces the observed profiles reasonably well and exhibits only moderate deviations which are discussed in the text. }
\end{figure}

\begin{figure}
\includegraphics[width=\columnwidth]{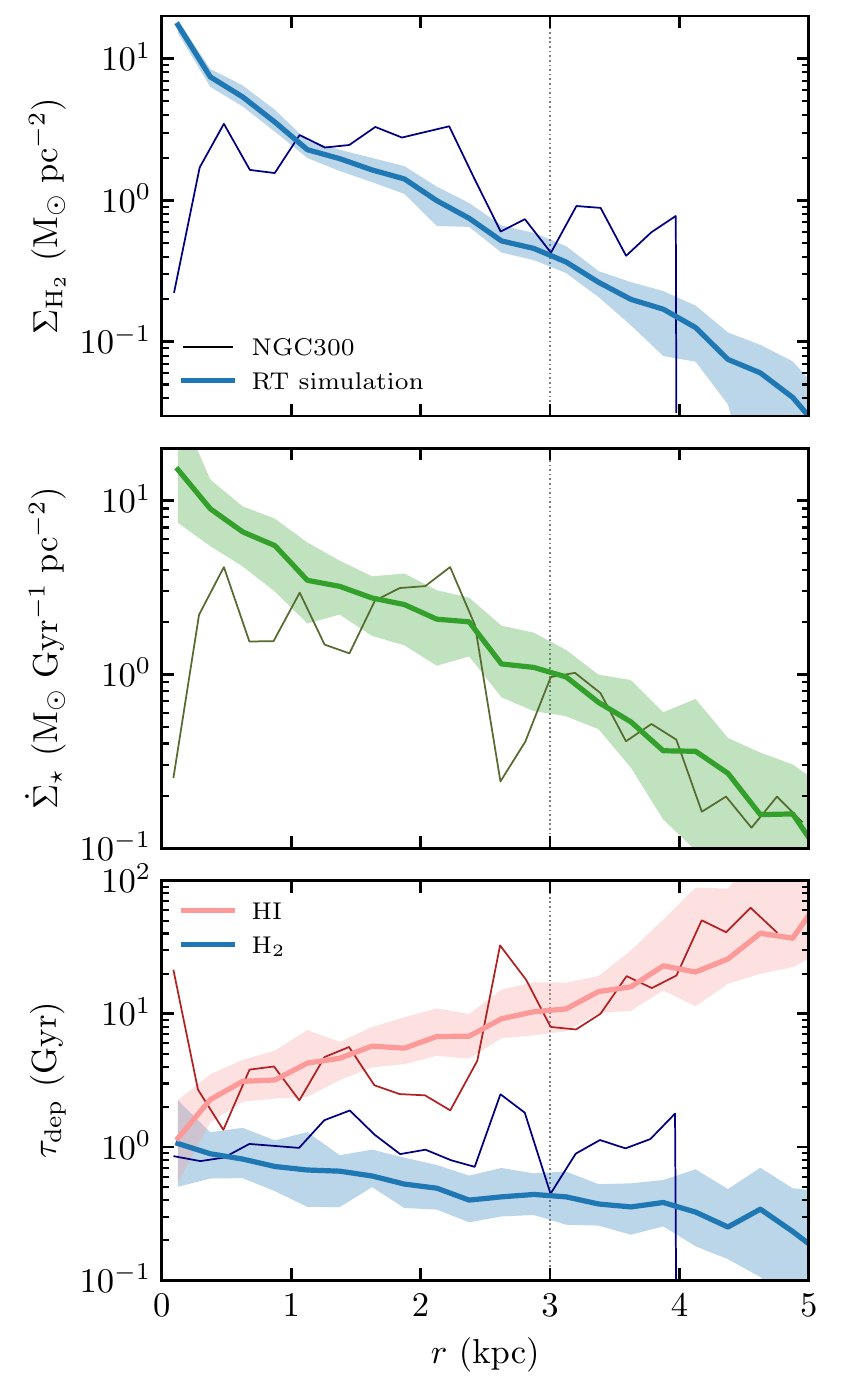}
\caption{\label{fig:prof-H2SFR} Comparison of the radial profiles for the key quantities used in our analysis: molecular gas (top panel) and SFR surface density (middle panel). The bottom panel also compares the depletion times of molecular (blue) and \ion{H}{1} gas (red). Thin lines show the profiles measured for NGC\,300 by \citet{Kruijssen.etal.2019}. Simulated profiles are stacked over 41 snapshots between 600 and 1000 Myr, with solid lines showing medians and the shaded region showing the snapshot-to-snapshot variation (16$^{\rm th}$--84$^{\rm th}$ interpercentile range). The vertical dotted line marks the region where we apply our analysis, $R<3\kpc$. Similar to the quantities shown in Figure~\ref{fig:prof-vcZ}, our fiducial simulation also reproduces the radial distribution of molecular gas and SFR reasonably well. }
\end{figure}

Figures~\ref{fig:maps}--\ref{fig:prof-H2SFR} overview the properties of the NGC\,300 analog from our fiducial simulation. 
Figure~\ref{fig:maps} shows a face-on view of the midplane slices of gas density, temperature, subgrid turbulent velocity, and the the radiation field at 12 eV (in the middle of the Lyman--Werner bands) computed by the RT solver and normalized to \citet{Draine.1978} units.
All these quantities exhibit orders of magnitude variations, and the overall morphology of the ISM is highly flocculent, reminiscent of the typical structure of sub-$L_\star$ galaxies \citep[e.g.,][]{Schombert.etal.1995}, including NGC\,300 \citep[see, e.g., the maps in Figure~1 of][]{Kruijssen.etal.2019}.

Figure~\ref{fig:prof-vcZ} shows that our galaxy remains structurally close to NGC\,300 over the time interval during which we carry out our analysis: radial profiles of rotational velocity, atomic gas and stellar surface densities, and metallicities all follow the NGC\,300 profiles reasonably well. The rotational velocities are slightly higher at $R<2\kpc$ due to a mild accumulation of dark matter mass near the disk center as a result of the initial relaxation.
Although the surface densities of total \ion{H}{1} may also seem to exceed the observed values, this difference is consistent with the contribution of the optically thick cold \ion{H}{1} that is missed in the observed $\SHI$ but can amount to $\sim$40\% of the total \ion{H}{1} mass in the Milky Way and nearby galaxies \citep[][]{Braun.etal.2009,Braun.2012}. To illustrate the possible contribution of cold \ion{H}{1}, the dashed red line shows only \ion{H}{1} gas warmer than 1000 K, which corresponds to $\sim$70\% of total \ion{H}{1} mass in our simulation.
Finally, the metallicity profile also exhibits a $\sim$30\% excess near the center due to continuous enrichment by stars formed during the simulation. 

Figure~\ref{fig:prof-H2SFR} shows the radial distribution of the quantities most relevant for our analysis: the surface densities of molecular gas, $\SH2$, and SFR, $\SSFR$. The bottom panel also shows the depletion time of molecular gas, $\tauH2 = \SH2/\SSFR$, and atomic gas, $\tauHI = \SHI/\SSFR$. All of these profiles match the corresponding profiles derived for NGC\,300 reasonably well. The agreement of the SFR profiles at $R>1\kpc$ is particularly remarkable because the star formation and feedback implementations were not calibrated for this galaxy and were also shown to work well in a more massive and metal-rich \Lstar~galaxy \citep{Semenov.etal.2017,Semenov.etal.2018,Semenov.etal.2019}. At $R<1\kpc$ the model SFR is somewhat larger than observed, which may be due to the elevated metallicity in that region and correspondingly enhanced cooling.
The $\tauH2$ profile is reasonably close to the observed near-constant value of $\sim1\Gyr$, although it does exhibit a slight negative trend with $R$ due to a somewhat steeper $\SH2$ profile predicted in the simulation. 

Overall, the differences between the model and observed profiles are rather modest, indicating that the global properties of our simulated galaxy are reasonably close to those of NGC\,300, enabling a direct comparison of the small-scale ISM structure. 

\begin{figure*}
\centering
{\large Effect of radiative feedback:}\\
\vspace{2pt}
\includegraphics[width=0.93\textwidth]{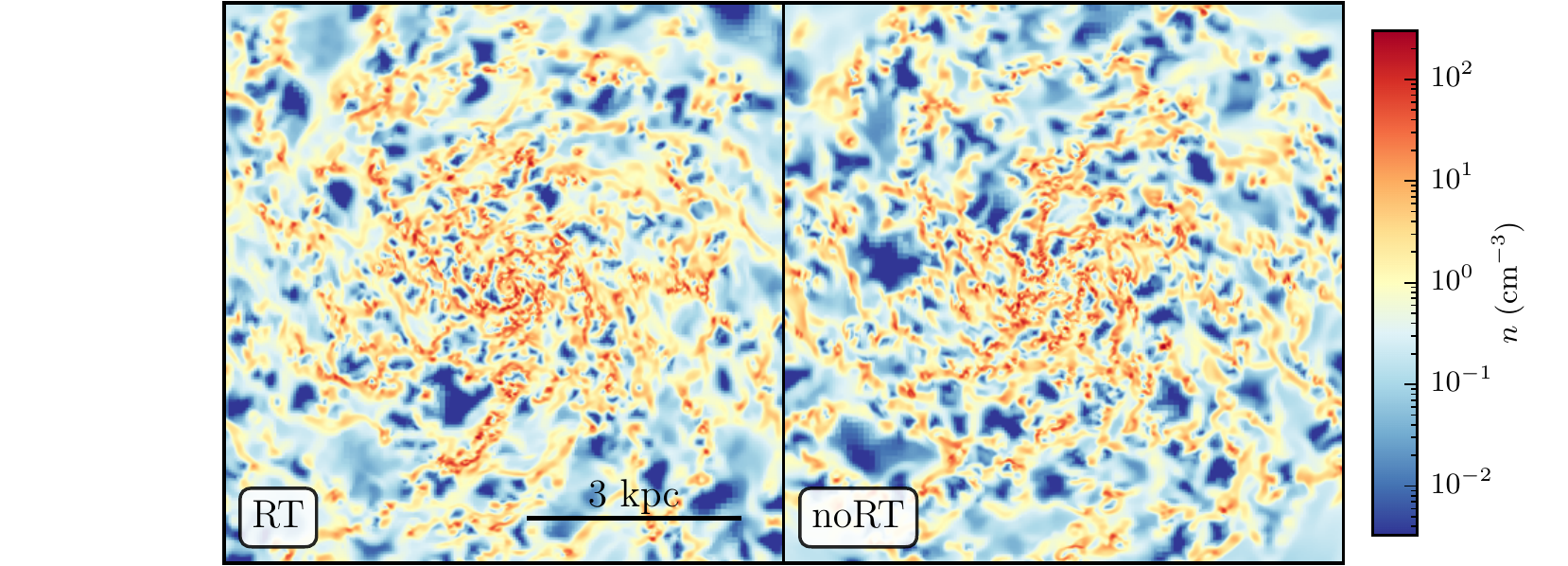}\\
\vspace{5pt}
{\large Effect of mechanical feedback:}\\
\vspace{2pt}
\includegraphics[width=\textwidth]{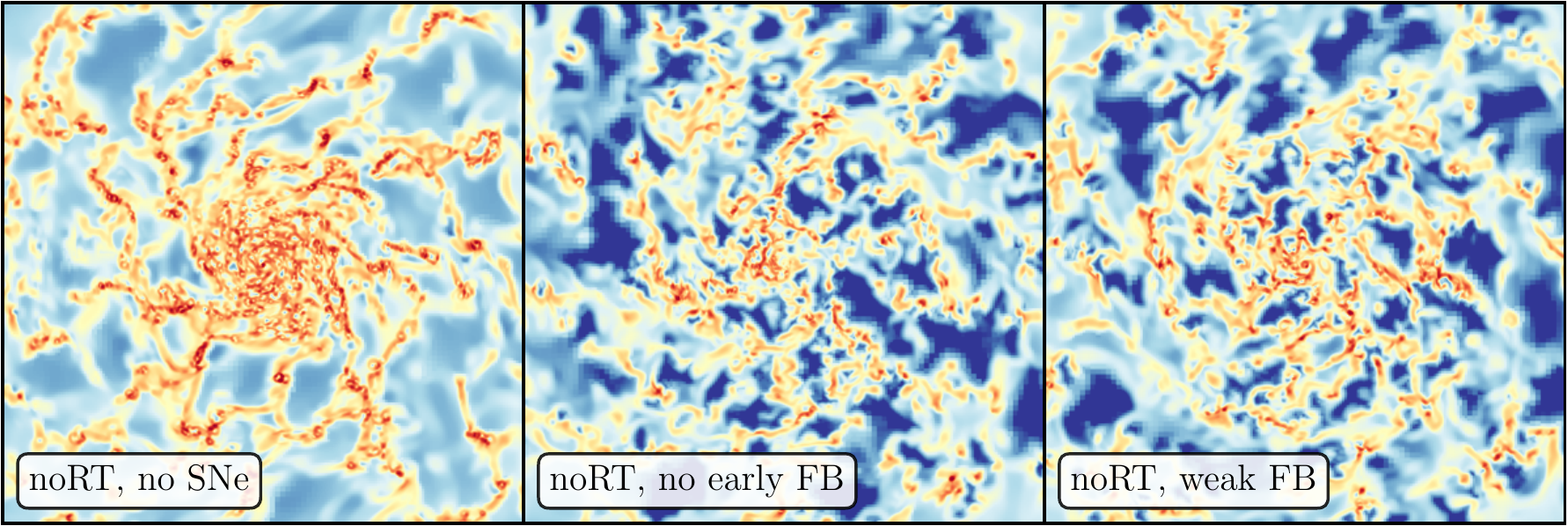}\\
\vspace{5pt}
{\large Effect of star formation prescription:}\\
\includegraphics[width=\textwidth]{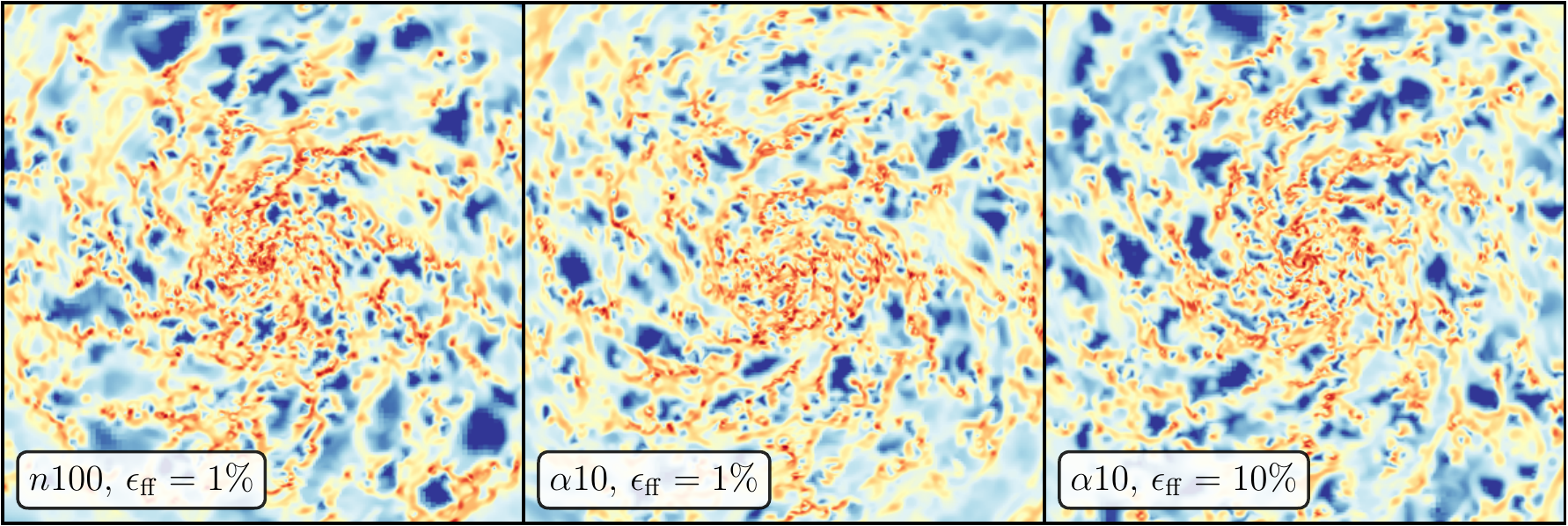}
\caption{\label{fig:maps-all} Midplane density slices from simulations with different feedback and star formation models explored in this paper. The simulation with an $\avir$ star formation threshold and $\epsff=3.3\%$ is qualitatively intermediate between the $1\%$ and $10\%$ cases and is not shown to save space. We checked that, in each of the runs, the ISM structure remains qualitatively similar in all available snapshots (e.g., see the animated version of Figure~\ref{fig:maps} in the online article), and therefore the maps demonstrate systematic differences between the models. In particular, variations of mechanical feedback strongly affect the global ISM structure, while changes in the star formation model result in a qualitatively similar structure. At the same time, as we will show below, both feedback and star formation models have a noticeable effect on the small-scale correlations between dense molecular gas and young stars. }
\end{figure*}

To give a visual impression of how different feedback and star formation models affect the global ISM structure, Figure~\ref{fig:maps-all} compares the midplane density slices of galaxies from our simulation suite. The most dramatic changes in the global gas structure are induced by variations of mechanical feedback (the second row of panels). For example, turning off SN feedback results in the ISM being devoid of tenuous hot bubbles (dark blue regions), with dense gas organized in prominent spiral structures, not typical for such a sub-\Lstar~galaxy. Also interesting, the models without early mechanical feedback and with a reduced momentum budget of feedback both result in qualitatively similar gas distributions. Dense regions in both simulations organize in more coherent $\sim$kpc-scale structures in contrast to the fiducial feedback model. This effect can be attributed to the reduced efficiency of star-forming gas dispersal and thus longer lifetimes of dense regions (see Section~\ref{sec:discussion:tsf} for further discussion).

Simulations with different star formation models, in contrast, all produce qualitatively similar global ISM structure. The structure of \emph{dense} gas, however, is significantly different (in particular, in the models with different $\epsff$), which leads to a strong effect on the correlation of dense gas and young stars as we will show in Section~\ref{sec:results:sf}.

We checked that, despite the variations of the ISM structure shown in Figure~\ref{fig:maps-all}, the bulk structural properties and radial profiles of quantities shown in Figures~\ref{fig:prof-vcZ} and \ref{fig:prof-H2SFR} remain reasonably close to NGC\,300 observations (the sensitivity of the $\SH2$, $\SSFR$, and $\tauH2$ profiles to feedback models is shown in Appendix~\ref{app:prof-varfb}). Therefore, we can investigate the effect of the star formation and feedback models on the correlation between dense gas and young stars without worrying about the global effects of the galaxy structure. 

\subsection{Tuning Fork Diagram Analysis}
\label{sec:sims:tf}

The key statistics that we explore in this paper is the tuning fork diagram \citep{Schruba.etal.2010,Kruijssen.Longmore.2014,Kruijssen.etal.2018,Kruijssen.etal.2019,Chevance.etal.2020}. As described in the Introduction, this diagram shows the relative bias of the depletion times measured in apertures of variable size placed on peaks of molecular gas or peaks in the distribution of young stars. Thus, to reproduce this observational statistic, the first step is to construct the maps of molecular gas and young stars and account for observational resolution and selection effects. 

To construct the molecular gas map, we project the volume density of H$_2$, $\rhoH2$, along the axis perpendicular to the disk plane. As detailed in Section~\ref{sec:sims:code}, in our simulations with RT, $\rhoH2$ is self-consistently computed in each cell using the six-species chemical network, while for runs without RT, we calibrated a model similar to \citet{Gnedin.Kravtsov.2011}. To mimic the sensitivity of CO observations of \citet{Kruijssen.etal.2019}, we apply two cuts to the resulting $\SH2$ maps: 
\begin{align}
\label{eq:SH2-min} \SH2 > \Sigma_{\rm H_2,min} = 13 \Msunpc2,\\
\label{eq:SH2-sigmamax} \sigma_{\rm turb}^{\rm proj} < 1\kms\left( \frac{\SH2}{\Sigma_{\rm H_2,min}} \right),
\end{align}
where $\sigma_{\rm turb}^{\rm proj}$ is the projected velocity dispersion in each pixel, which includes the contribution of subgrid turbulence (see Section~\ref{sec:sims:code}) and resolved velocity dispersion along the line of sight.  
The second cut approximates the loss of sensitivity due to the increasing width of the CO line \citep[e.g.,][]{Sun.etal.2018} and possible dependence of the CO-to-H$_2$ conversion factor on local turbulence. The parameters in this cut were chosen to qualitatively reproduce the CO map in NGC\,300 from \citet{Kruijssen.etal.2019} by removing highly turbulent moderate-density molecular regions and extended outskirts of gas peaks that are not present in the observed map.
As we show in Appendix~\ref{app:map-cuts}, the opening of the tuning fork is quite sensitive to the specific choice of the cuts (see Figure~\ref{fig:tf-var-H2-cuts}), and we discuss this issue further in Section~\ref{sec:discussion}.

To construct a map of recent SFR, we select young star particles with ages $2\text{--}5\Myr$ and compute $\SSFR$ in each pixel as $M_{\star,2\text{--}5\Myr}/\Delta t_{\rm SFR}/({\rm pixel\ area})$, where $\Delta t_{\rm SFR} = 3\Myr$. The lower age cut approximates the typical observational estimates for the duration of the embedded star formation stage of $\approx 2\text{--}3$ Myr \citep[e.g.,][]{Lada.Lada.2003,Corbelli.etal.2017,Kim.etal.2020}. The upper age cut corresponds to the timescale over which young stellar population is expected to be seen in \Halpha~\citep{Kennicutt.Evans.2012,Haydon.etal.2020,FloresVelazquez.etal.2021}. The effect of these age cuts on the tuning fork is shown in Appendix~\ref{app:map-cuts}.

To mimic the analysis of \citet{Kruijssen.etal.2019} we smooth our $\SH2$ and $\SSFR$ maps using a 2D Gaussian filter with a width of 20 pc and use only the inner $R<3\kpc$ for our analysis. We also tried excluding the central $1\kpc$ where some of the radial profiles deviate from those derived for NGC\,300 (see Section~\ref{sec:sims:ics}), but the effect of such exclusion on the tuning fork is small.

To identify gas and SFR peaks, we used the local extrema of the $\SH2$ and $\SSFR$ maps. \citet{Kruijssen.etal.2019} use a more complex algorithm \citep[\textsc{Clumpfind};][]{clumpfind} that can extract peaks from the noisy observed maps. Our simulated maps, however, are sufficiently smooth so that a simple method is sufficient.

Finally, to construct the tuning fork diagram we followed the steps outlined by \citet{Kruijssen.etal.2018}. For a given scale $L$, we smooth the maps with a top hat filter with the window size $L$ and compute the average $\SH2$ and $\SSFR$ at the locations of gas and SFR peaks identified above. When $L$ is sufficiently large for some of the apertures to overlap, we randomly subsample nonoverlapping peaks and average resulting $\SH2$ and $\SSFR$ over 100 such Monte Carlo samples. Thus, the depletion time $\tauH2=\SH2/\SSFR$ in apertures of size $L$ centered on a given peak type are computed from $\SH2$ and $\SSFR$ that are averaged over the peaks in each Monte Carlo sample and over all samples. 

Repeating this procedure for different $L$ produces the tuning fork diagram for a single snapshot. To reduce the noise due to snapshot-to-snapshot variation, we compute the diagram for all snapshots available in the simulation (see Section~\ref{sec:sims:ics}) and show the median $\tauH2$ at each $L$. To estimate the magnitude of the snapshot-to-snapshot variation we also compute 2.5, 16, 84, and 97.5 percentiles, which approximate the boundaries of 1$\sigma$ and 2$\sigma$ deviation for a Gaussian distribution.

\section{Results}
\label{sec:results}

\subsection{Tuning Fork Diagram in the Fiducial Simulation}
\label{sec:results:tf-fid}

\begin{figure}
\includegraphics[width=\columnwidth]{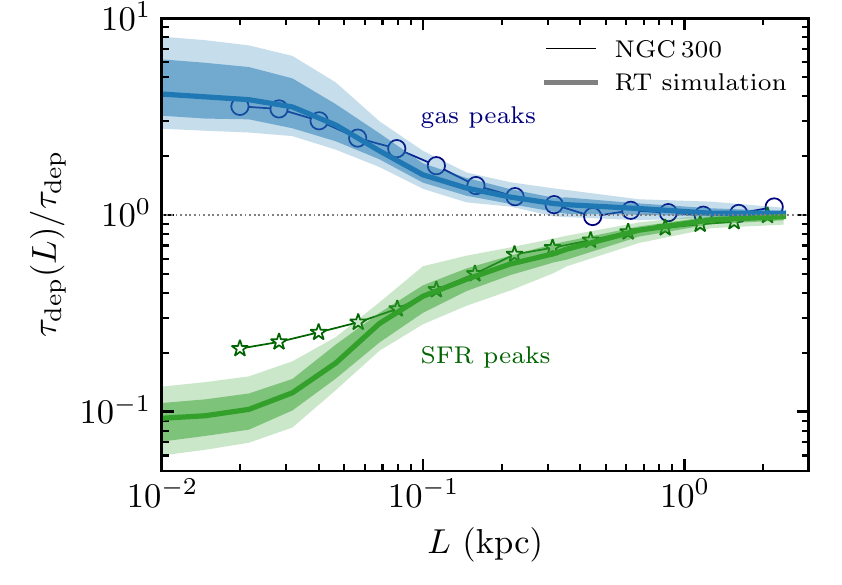}
\caption{\label{fig:tf-rt} Comparison of the tuning fork diagram from our fiducial simulation---thick lines and shaded regions---with the measurements of \citet{Kruijssen.etal.2019}---thin lines with markers. The diagram reflects the deviation of $\tauH2$ from the galaxy-averaged value ($y$-axis) in apertures of variable size $L$ ($x$-axis) centered on H$_2$ (upper branch) or recent SFR peaks (lower branch; see Section~\ref{sec:sims:tf} for details). Simulation results are stacked over 41 snapshots between 600 and 1000 Myr, with the thick lines showing the medians and shaded regions showing 1$\sigma$ and 2$\sigma$ of the snapshot-to-snapshot variation. The simulation can reproduce the observed wide opening of the tuning fork, indicating that our fiducial star formation and feedback model is reasonably realistic.}
\end{figure}

We start from the comparison of the tuning fork diagram from our fiducial simulation with the measurements for NGC\,300 by \citet{Kruijssen.etal.2019} shown in Figure~\ref{fig:tf-rt}. The diagram is computed as described in Section~\ref{sec:sims:tf}. The solid lines show the median relation, and shaded regions indicate the snapshot-to-snapshot variation.

As the figure shows, our fiducial simulation reproduces the observed tuning fork-like shape remarkably well, especially at $L \gtrsim 100\pc$.
At large scales, $L \gtrsim 1\kpc$, the branches of the fork converge as both types of apertures include a large number of gas and SFR peaks and their $\tauH2$ approach the galaxy-averaged value. At smaller scales, the branches diverge, indicating the preference for sampling of either non-star-forming gas or SFR regions with little molecular gas by corresponding apertures. The divergence scale corresponds to the average separation between the peaks of molecular gas and star formation. As the figure demonstrates, this scale is reproduced remarkably well. Finally, the wide opening of the tuning fork at the smallest $L$ reflects the level of spatial decorrelation between recent SFR events and dense gas on these scales. 

The opening of the tuning fork at $L \lesssim 50\pc$ in our simulation is somewhat larger than observed. These scales approach the resolution of our simulation ($\Delta = 10\pc$), which hinders the interpretation of the results on these scales. However, it is worth noting that our simulations and analysis do not include some of the factors that can increase the correlation between molecular gas and SFR and thus reduce the tuning fork opening. For example, a more realistic modeling of \Halpha~emission, instead of using young star particles as a proxy, can increase the size of the SFR peaks, leading to a stronger correlation with $\SH2$ on small scales (see Section~\ref{sec:discussion} for further discussion). 
Nevertheless, the achieved qualitative and quantitative agreement is rather remarkable, as recent galaxy simulations struggled to reproduce the opening of the tuning fork even on $>50\pc$ scales \citep[e.g.,][]{Fujimoto.etal.2019}. 

As our fiducial simulation can reproduce the observed tuning fork reasonably well, we can now compare our simulations with different models of feedback and star formation to investigate the relative role of various processes in determining the opening of the tuning fork and check whether this statistics can be used to constrain such models. The snapshot-to-snapshot variation of the tuning fork, shown with the shaded regions in Figure~\ref{fig:tf-rt}, can be used as a reference in this comparison.

\subsection{Variation of the Feedback Model}
\label{sec:results:fb}

\begin{figure}
\includegraphics[width=\columnwidth]{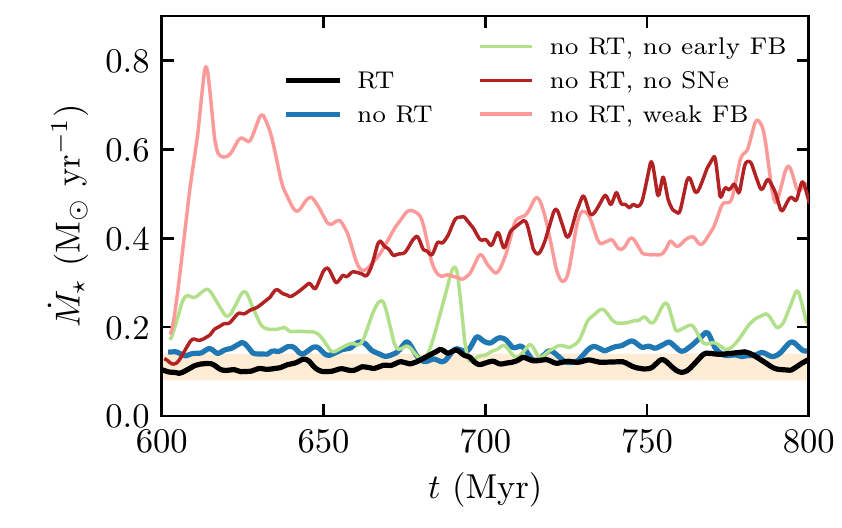}
\caption{\label{fig:sfh-varfb} SFR histories from our NGC\,300 simulations with variations of the feedback model. The orange stripe indicates the range of global SFR values estimated for NGC\,300: $\SFR \sim 0.08\text{--}0.14\Msunyr$ \citep{Helou.etal.2004}. Our fiducial RT run, and the runs without RT and without early mechanical feedback all produce global SFR close to NGC\,300 observations, while the runs without SNe and with feedback momentum reduced by a factor of 5 lead to a $\sim3\text{--}4$ times higher SFR.}
\end{figure}

\begin{figure}
\includegraphics[width=\columnwidth]{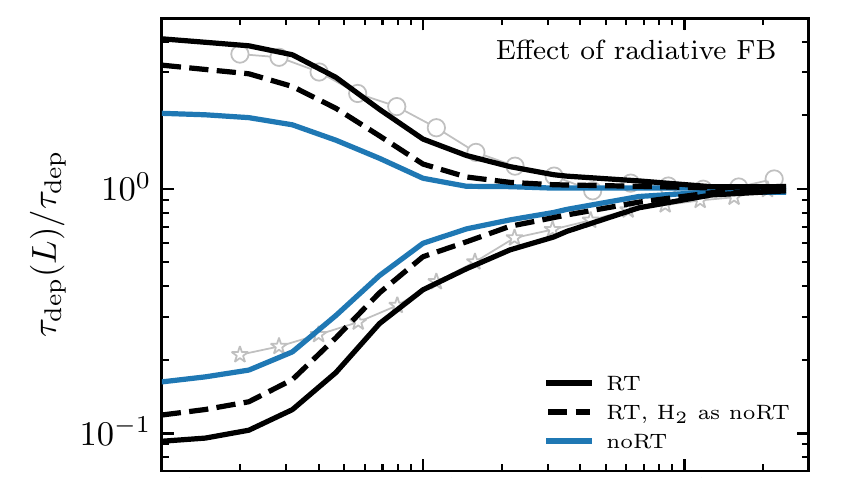}\\
\includegraphics[width=\columnwidth]{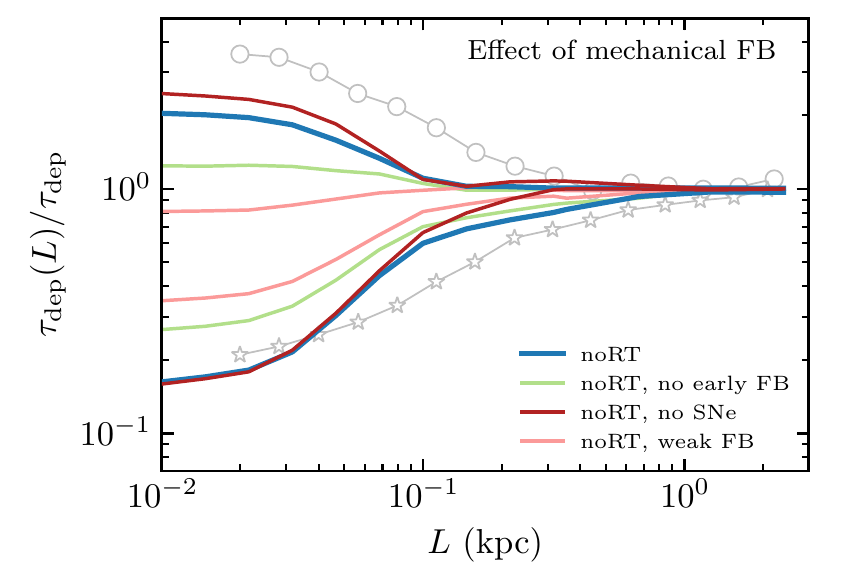}
\caption{\label{fig:tf-varfb} Effect of the feedback model on the tuning fork diagram. Line colors repeat those in Figure~\ref{fig:sfh-varfb}. The top and bottom panels show the effect of explicit RT and H$_2$ chemistry modeling and the effect of mechanical feedback variation, respectively. To demonstrate the effect of H$_2$ modeling, the dashed black lines show the results of our fiducial RT simulation, where $\rhoH2$ is computed using the model calibrated for runs without RT (see Section~\ref{sec:sims:code}). Simulation results are stacked over available snapshots, and lines show the median values (see Section~\ref{sec:sims:tf}). For reference, thin gray lines with markers show the measurements in NGC\,300 reported by \citet{Kruijssen.etal.2019}. Each of the explored feedback processes has a noticeable effect on the diagram, but neither of the processes dominate. The wide opening of the tuning fork results from a combined effect of multiple different processes.}
\end{figure}

\begin{figure*}
\includegraphics[width=\textwidth]{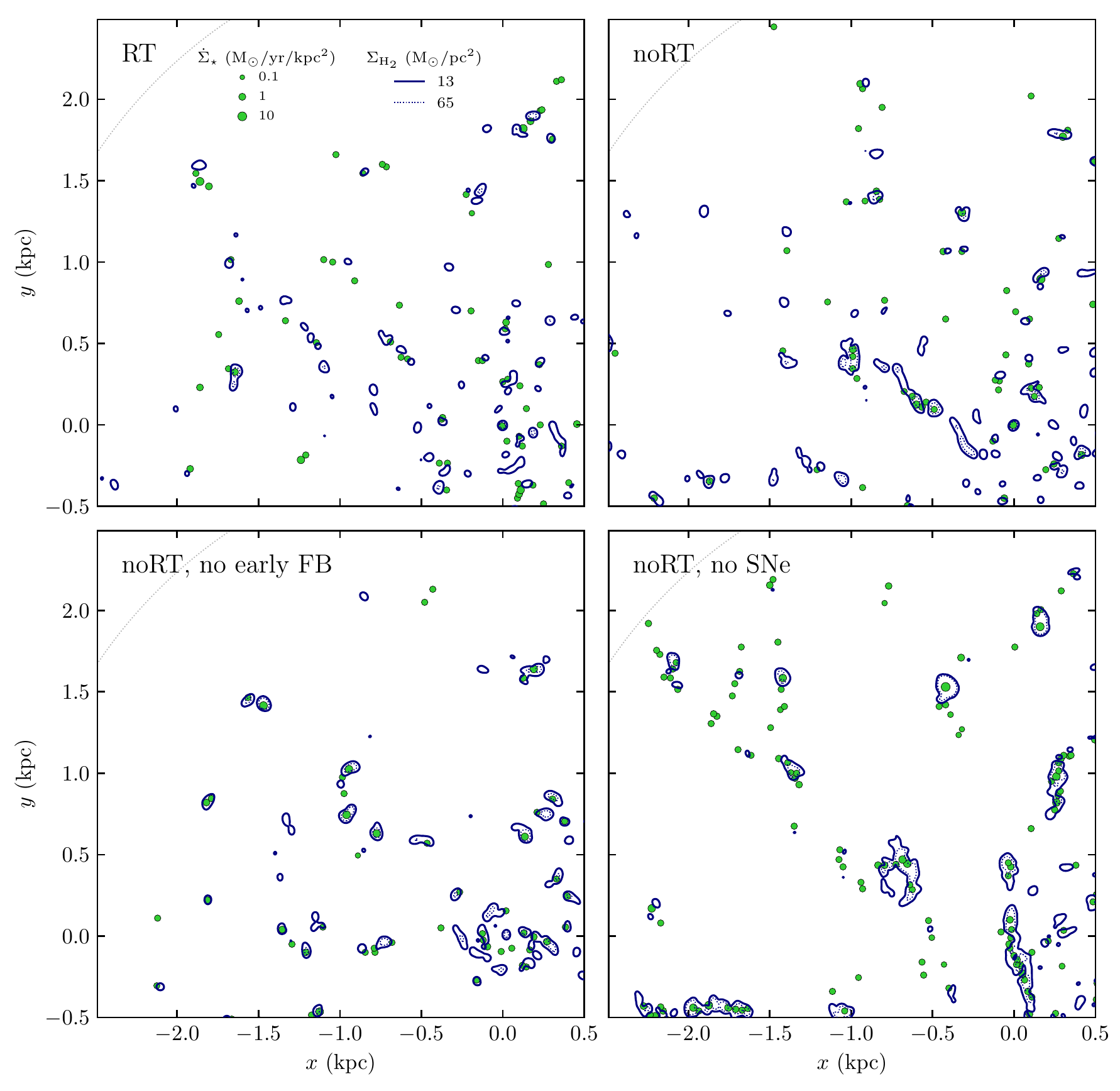}
\caption{\label{fig:peakmaps-varfb} Spatial distribution of molecular gas (blue contours) and recent SFR peaks (green circles) used to compute the tuning fork diagram in our simulations with variations of the feedback model. The shown region is a zoom-in to the top left quadrant of the galaxy, with the galaxy center at $(x,y)=(0,0)$. The $\SSFR$ map is computed using the star particles with ages 2--5 Myr (see Section~\ref{sec:sims:tf}), and green circles indicate the locations of $\SSFR$ peaks, with sizes changing logarithmically with the SFR value at the peak location as shown in the legend in the top left panel. The dotted arc in the top left corner of each panel marks the boundary of the region used for analysis, $R<3\kpc$. As we turn off RT and then early feedback, gas peaks become more prominent and more strongly correlated with recent SFR peaks, leading to a narrowing down of the tuning fork diagram in Figure~\ref{fig:tf-varfb}. Turning off SNe leads to the largest gas peaks but weak correlation with young stars on small scales. As a result, the tuning fork in this run is very close to the ``noRT'' case at $<100 \pc$ scales but much narrower on larger scales. }
\end{figure*}

To investigate the role of different feedback processes in shaping the tuning fork, we ran a series of models using a uniform UV background instead of self-consistent RT and adopting different assumptions about the momentum injection from young star. Specifically, we explored uniform UV (no RT) simulations with (i) the same feedback model as in the fiducial run, (ii) the feedback model without early feedback, (iii) the model without SNe, and (iv) the model with both pre-SN and SN feedback but with a reduced momentum injection rate by a factor of 5. These variations enable us to test separate effects of radiative feedback, pre-SN momentum injection, and the strength of SN feedback on the tuning fork diagram. In addition, to gauge the effect of the H$_2$ model, we reanalyzed our fiducial RT simulations, disregarding the results of molecular chemistry calculations and using the approximate model calibrated for use in the runs without RT (Section~\ref{sec:sims:code}).

To ensure a fair comparison, we checked that the radial profiles of $\SH2$ and $\SSFR$ remain similar in all of the resimulations (see Appendix~\ref{app:prof-varfb}). The only exception are the runs with weak feedback and without SNe, in which $\SSFR$ and global SFR increase by a factor of three to five consistent with a strong sensitivity of the SFR to the feedback energy and momentum budget found in many other simulations  \citep[e.g.,][]{Agertz.Kravtsov.2015,Hopkins.etal.2017,Orr.etal.2017,Semenov.etal.2018}. As a result, the global SFR in these runs becomes inconsistent with NGC\,300 observations as demonstrated in Figure~\ref{fig:sfh-varfb}. In addition, the run without SNe also produces a factor of $\sim$5 more H$_2$. Nevertheless, it is still interesting to investigate the effect of such feedback models on the tuning fork. 

Tuning fork diagrams from these simulations are compared in Figure~\ref{fig:tf-varfb}. The differences between the cases illustrate the effects of each feedback process, as described below. In addition, Figure~\ref{fig:peakmaps-varfb} provides a visual illustration of the differences in  the distribution of dense gas and recent SFR peaks used in the analysis. 

\emph{Photodissociation of H$_2$.} The dashed black line in the top panel of Figure~\ref{fig:tf-varfb} shows the results of our fiducial RT simulation, which we reanalyzed using the same H$_2$ model as in the runs without RT instead of the on-the-fly molecular chemistry calculations (the latter is shown with the solid black line). This H$_2$ model is calibrated to reproduce the radial profile of $\SH2$ within $R<3\kpc$ from the full RT simulation by selecting an effective average UV field instead of using the local value (see Section~\ref{sec:sims}). Thus, the difference between the solid and dashed black lines shows the effect of H$_2$ photodissociation by the spatially variable UV field.

The figure shows that self-consistent treatment of H$_2$ formation and dissociation does have some effect. This effect, however, is modest and is comparable to the effects of other feedback and star formation processes considered below.  

\emph{Spatially nonuniform gas heating.} The solid blue line shows the tuning fork diagram in the calculation without RT modeling, assuming a constant background UV field instead. Thus, the difference with the dashed black line shows the effect of a self-consistently computed UV field on the local heating rate of the star-forming gas. As the top right panel in Figure~\ref{fig:peakmaps-varfb} shows, molecular gas peaks become more prominent and more strongly correlate with the locations of young stars, resulting in a somewhat smaller opening of the tuning fork. Here again the effect is modest and comparable to the effects of other variations.

\emph{Early mechanical feedback.} The green line in the bottom panel shows the case where we turn off our approximate model for early feedback and allow star-forming regions to accumulate gas unimpeded during the first 3 Myr after the onset of star formation, before the first SN explosion.
Interestingly, this makes the overall ISM structure less flocculent (see Figure~\ref{fig:maps-all}), but the effect on the tuning fork diagram opening is nevertheless small, as can be seen from the comparison of the blue and green lines. This is because, despite the differences in the global ISM structure, the distribution of \emph{dense} gas and recent SFR peaks in these two runs is qualitatively similar, with a stronger correlation between the peaks in the run without early feedback (see the top right and bottom left panels in Figure~\ref{fig:peakmaps-varfb}).

\emph{Type II SN feedback.} Next, the dark red line in the bottom panel of Figure~\ref{fig:tf-varfb} shows the simulation where the SNe are turned off and young stars inject energy and momentum only during the first 3 Myr after the onset of star formation. The opening of the tuning fork at small $L$ is very close to that in the simulation with SNe (blue line), while on $\gtrsim 100\pc$ scales the opening is strongly reduced. The similarity of the tuning fork at small $L$ indicates that the correlation between gas and young stars on these scales is set by efficient early feedback. At the same time, SNe strongly affect the ISM structure on $\gtrsim 100\pc$ scales as is also clear from Figure~\ref{fig:maps-all} and the bottom right panel of Figure~\ref{fig:peakmaps-varfb}.
Such a sensitivity of the tuning fork diagram on different $L$ to different feedback processes is consistent with the results of \citet{Li.etal.2020}, who found a qualitatively similar behavior of the two-point correlation function of giant molecular clouds. 

\emph{Overall feedback strength.} Finally, the pale red line in Figure~\ref{fig:tf-varfb} shows the simulation where both pre-SN and SN feedbacks operate but the overall momentum injection rate of feedback is reduced by a factor of five. Interestingly, this has a stronger effect on the tuning fork opening than turning off pre-SN feedback and keeping fiducial SN momentum (compare blue and pale red lines). This result indicates that SNe can contribute to the tuning fork opening at small $L$ when early feedback is weak and does not dominate the effect. Indeed, early feedback is stronger in the ``weak FB'' case, while SNe are stronger in the ``no early FB'' case, and the latter run results in a wider tuning fork opening as is clear from the comparison of the pale red and green lines in the figure.

All in all, Figure~\ref{fig:tf-varfb} shows that each of the factors explored above gradually reduces the opening of the tuning fork, but their individual effects are modest and comparable to the snapshot-to-snapshot variation (see Figure~\ref{fig:tf-rt}). \emph{Thus, the degree of the tuning fork opening comparable to that measured in observations results from a combined effect of multiple feedback aspects.} 

Apart from stellar feedback, many other factors have a comparable or even stronger effect on the tuning fork opening. For example, the assumptions about the embedded stage of star formation, the timescale over which young stars produce \Halpha~emission, and the selection effects of molecular gas observations all strongly affect the tuning fork opening as detailed in Appendix~\ref{app:map-cuts}.

\subsection{Variation of the Star Formation Prescription}
\label{sec:results:sf}

\begin{figure}
\includegraphics[width=\columnwidth]{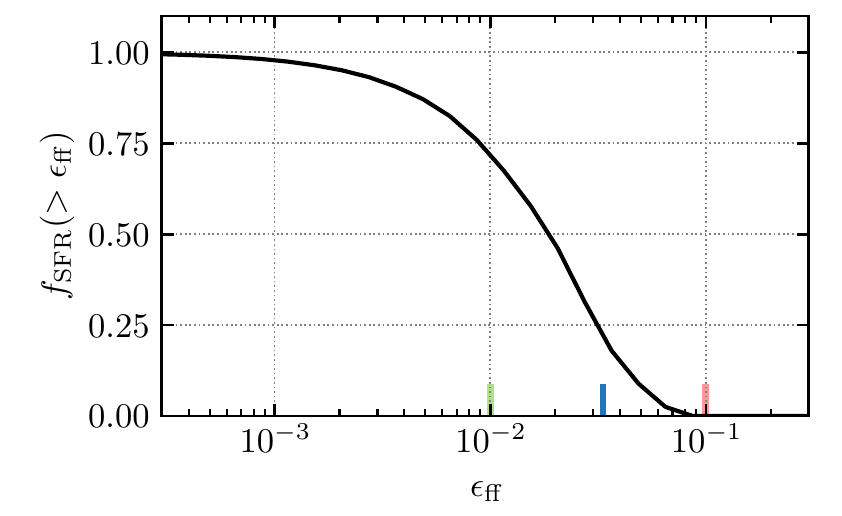}
\caption{\label{fig:epsff} SFR-weighted cumulative distribution of $\epsff$ in our fiducial simulation, where $\epsff$ varies exponentially with local $\avir$ using the \citet{Padoan.etal.2012} model. The distribution of $\epsff$ in actively star-forming cells is broad, ranging between $\epsff\sim0.1\%$ and $10\%$. Vertical colored ticks show the fixed $\epsff$ values used in our tests with a star formation threshold in $\avir$: $\epsff=1\%$, $3.3\%$, and $10\%$. The cells with $\epsff>1\%$ and $\epsff>3.3\%$ contribute $\approx 70\%$ and $20\%$ of total SFR, respectively, while $\epsff=10\%$ corresponds to the highest $\epsff$ value reached in the simulation. }
\end{figure}

\begin{figure}
\includegraphics[width=\columnwidth]{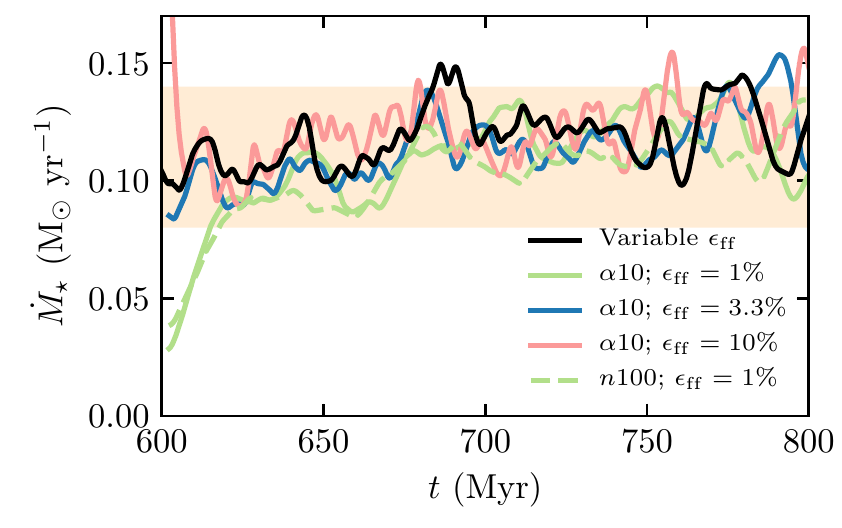}
\caption{\label{fig:sfh-varsf} SFR histories in our simulations with different star formation prescriptions, i.e., different assumptions about local $\epsff$ and star formation threshold (see Section~\ref{sec:sims:code} and Table~\ref{tab:sims}). As in Figure~\ref{fig:sfh-varfb}, the orange stripe indicates the range of global SFR values estimated for NGC\,300: $\SFR \sim 0.08\text{--}0.14\Msunyr$ \citep{Helou.etal.2004}. All explored variations of the star formation model result in approximately the same global SFR. In particular, the global SFR is insensitive to the local $\epsff$ value, implying that our simulated NGC\,300 analog is in the self-regulated regime. }
\end{figure}

\begin{figure}
\includegraphics[width=\columnwidth]{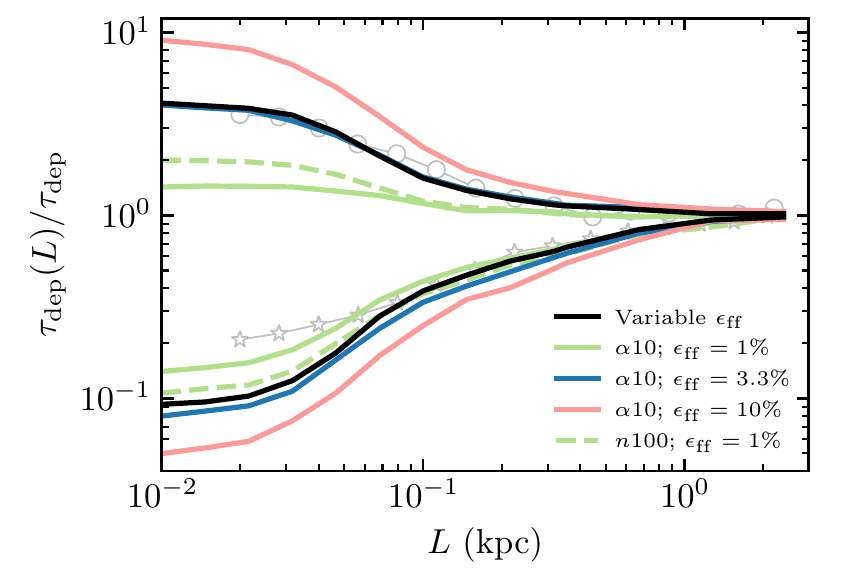}
\caption{\label{fig:tf-varsf} Effect of the star formation model on the tuning fork diagram. Line colors and styles repeat those in Figure~\ref{fig:sfh-varsf}. Simulation results are stacked over available snapshots, and lines show the median values (see Section~\ref{sec:sims:tf}). For reference, thin gray lines with markers show the observed relation in NGC\,300 from \citet{Kruijssen.etal.2019}. The opening of the tuning fork strongly depends on the choice of $\epsff$, even though the global SFR is almost insensitive to its value (see Figure~\ref{fig:sfh-varsf}). To reproduce the tuning fork in our fiducial model with variable $\epsff$, a relatively high value of constant $\epsff=3.3\%$ is needed---gas with such high $\epsff$ accounts for only $\approx20\%$ of the total SFR in the fiducial simulation (see Figure~\ref{fig:epsff}). At the same time, the models with $\epsff=1\%$ and $10\%$ under- and overpredict the tuning fork opening, respectively. The choice of the star formation threshold has a relatively minor effect for these simulations.}
\end{figure}

\begin{figure*}
\includegraphics[width=\textwidth]{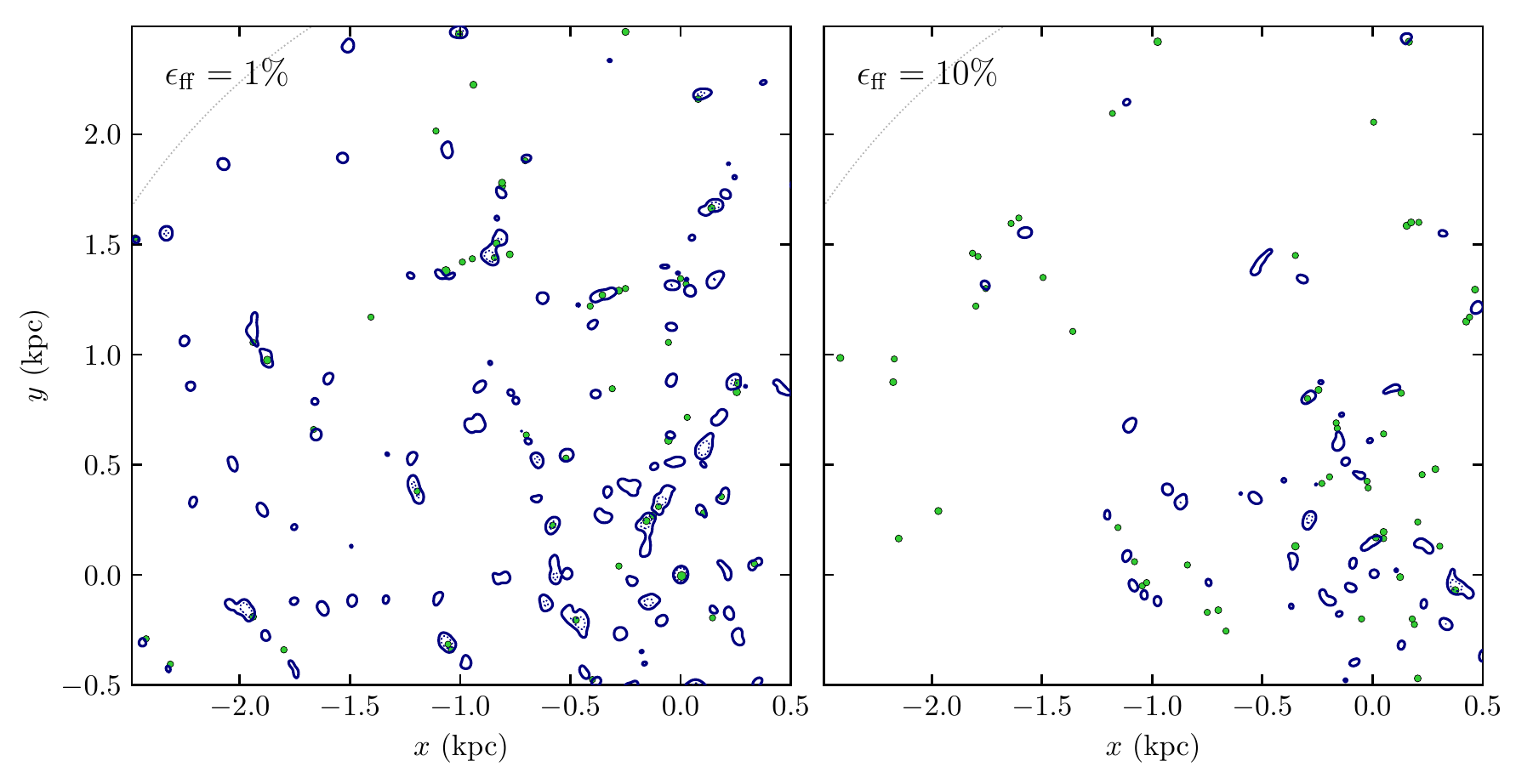}
\caption{\label{fig:peakmaps-varsf} Same as Figure~\ref{fig:peakmaps-varfb} but for the simulations with the star formation threshold $\avirsf=10$ and different values of $\epsff$: $1\%$ and $10\%$. The figure shows that at higher $\epsff$, the instantaneous amount of molecular gas (blue contours) strongly decreases, leading to a weaker correlation with the peaks of recent SFR (green circles) and a wider opening of the tuning fork diagram. Such a dramatic effect on molecular gas fraction is due to increased efficiency of star-forming gas dispersal at higher $\epsff$, which is consistent with the picture of SFR self-regulation via rapid gas cycling in the ISM as detailed in the Discussion section. }
\end{figure*}

Another major factor that can affect the tuning fork in galaxy simulations is the star formation prescription, e.g., the choice of the local star formation efficiency per freefall time ($\epsff$, see Equation~\ref{eq:rhosfr}) and the criteria used to select star-forming gas. Many recent galaxy simulations showed that when stellar feedback is efficient, global SFR and depletion time can become insensitive to the choice of local $\epsff$ \citep[e.g.,][]{Dobbs.etal.2011,Agertz.etal.2013,Hopkins.etal.2013,Hopkins.etal.2017,Orr.etal.2017}, which can be explained by the efficient dispersal of star-forming regions by feedback \citep{Semenov.etal.2017,Semenov.etal.2018}. 
The  possible dependence of the tuning fork on star formation parameters is therefore particularly interesting as it can help to constrain these parameters in such a self-regulated regime.

In our fiducial simulation, the value of $\epsff$ varies continuously as an exponential function of the local virial parameter (Equation~(\ref{eq:epsff-P12})). The cumulative distribution of $\epsff$ weighted by local SFR is shown in Figure~\ref{fig:epsff}. The values of $\epsff$ in actively star-forming regions in this run range between $\sim0.1\%$ and $10\%$, with half of the total SFR produced in cells with $\epsff>2\%$.
The average $\langle \epsff \rangle$ depends on weighting, and for the averages weighted by $\rhoSFR$, $\rho/\tff$,\footnote{The averaging by $\rho/\tff$ is motivated by Equation~(\ref{eq:rhosfr}), and for a given distribution of $\rho$ it produces the same instantaneous total SFR as the variable $\epsff$.} and $\rho_{\rm H_2}$ we find $\approx 2.2\%$, $0.9\%$, and $1.0\%$, respectively. To avoid including non-star-forming gas in the averages, we select only the cells with $\epsff > 0.1\%$, which account for $\approx98\%$ of global SFR in our fiducial simulation. 

To explore the effect of the $\epsff$ choice, we also rerun our RT simulation where the exponential dependence of $\epsff$ on $\avir$ is approximated by an $\avir$ threshold and a fixed value of $\epsff$ in gas with $\avir<\avirsf$. For our tests, we adopt $\avirsf=10$ and explore cases of $\epsff=1\%$, $3.3\%$, and $10\%$ that probe the range of $\epsff$ values realized in our fiducial simulation (see the colored ticks in Figure~\ref{fig:epsff}). 
To explore the effect of the star formation threshold, we also reran our simulation with a threshold in density, $n>\nsf=100\cc$, that selects approximately the same amount of gas as being star-forming as our $\avir$ threshold.

As Figure~\ref{fig:sfh-varsf} shows, the global SFR is insensitive to the changes in the star formation prescription that we explored and remains consistent with the NGC\,300 observations. In particular, the global SFR is insensitive to a 10-fold change of $\epsff$, implying that our simulated NGC\,300 analog is in the self-regulated regime.

In contrast, the tuning fork diagram depends strongly on the adopted $\epsff$ value as shown in Figure~\ref{fig:tf-varsf}: the opening of the tuning fork increases with increasing $\epsff$ and becomes wider than the observed diagram for $\epsff$ significantly larger than $3\%$. Thus, the observed tuning fork diagram in NGC\,300 prefers the small $\epsff$ values of a few percent consistent with the observational estimates in the Milky Way and nearby star-forming galaxies \citep[e.g.,][]{Krumholz.Tan.2007,Lee.etal.2016}. These results also echo our previous findings about the dependence of the tuning fork on the star formation prescription in simulations of an \Lstar-sized galaxy \citep[see Figure~11 in][]{Semenov.etal.2018}.

Interestingly, the tuning fork diagram in our fiducial run can be closely reproduced in runs with the $\avir$ star formation threshold, but only by setting a fairly high $\epsff$ value of 3.3\%. As can be seen in Figure~\ref{fig:epsff}, cells with such high $\epsff$ contribute only $\approx 20\%$ of the total SFR. At the same time, $\epsff=1\%$, which is close to the average $\langle \epsff \rangle$ weighted by $\rho/\tff$ or $\rho_{\rm H_2}$, produces a smaller opening of the tuning fork diagram than observed. This result demonstrates that the opening of the diagram is sensitive to the star-forming regions in the tail of large values in the $\epsff$ distribution, with the effective $\epsff$ value closer to the $\langle \epsff \rangle$ weighted by local SFR ($\approx 2.2\%$ in our fiducial simulation).

The origin of the effect of $\epsff$ on the tuning fork diagram is clear from Figure~\ref{fig:peakmaps-varsf}, which shows the distribution of molecular gas and young stars from the simulations with fixed $\epsff=1\%$ and $10\%$. In the run with $\epsff=10\%$, the number of young stars formed over the $\sim10\Myr$ interval (and thus the global SFR) is the same as in the run with $\epsff=1\%$, but the instantaneous number of gas peaks is much smaller in the run with larger $\epsff$. As a result, the correlation between gas and SFR peaks becomes weaker, leading to a wider opening of the tuning fork diagram. The physical origin of these effects is due to the strong decrease of star-forming gas lifetimes at high $\epsff$ values, as we further discuss in Section~\ref{sec:discussion:cycle}. 

Apart from the effect of $\epsff$, Figure~\ref{fig:tf-varsf} also shows the dependence of the tuning fork on the choice of the star-formation threshold either in $\avir$ or density. The choice of the threshold may change the correlation between molecular gas and young stars by changing the correlation between molecular gas and star-forming regions. However, as the figure shows, the magnitude of this effect is small. Note, however, that the effect magnitude can depend on the global properties of the galaxy. For example, as was shown in \citet{Semenov.etal.2019}, in an \Lstar~galaxy, the choice of the star formation threshold can lead to a qualitatively different correlation between molecular gas and SFR even on a $\sim$kiloparsec scale.

Finally, as the simulation with the density-based star formation threshold  does not require modeling of subgrid turbulence, we also tested the effect of turbulence on the tuning fork by switching it off in this simulation. We find, however, that this run produces the global SFR and the tuning fork diagram very close to the results of the run with the subgrid turbulence modeling, indicating that the \emph{dynamical} effect of unresolved turbulent pressure is small.

\section{Discussion}
\label{sec:discussion}

\subsection{Sensitivity of the Tuning Fork Diagram to the Star Formation--Feedback Cycle}
\label{sec:discussion:cycle}

\begin{figure}
\includegraphics[width=\columnwidth]{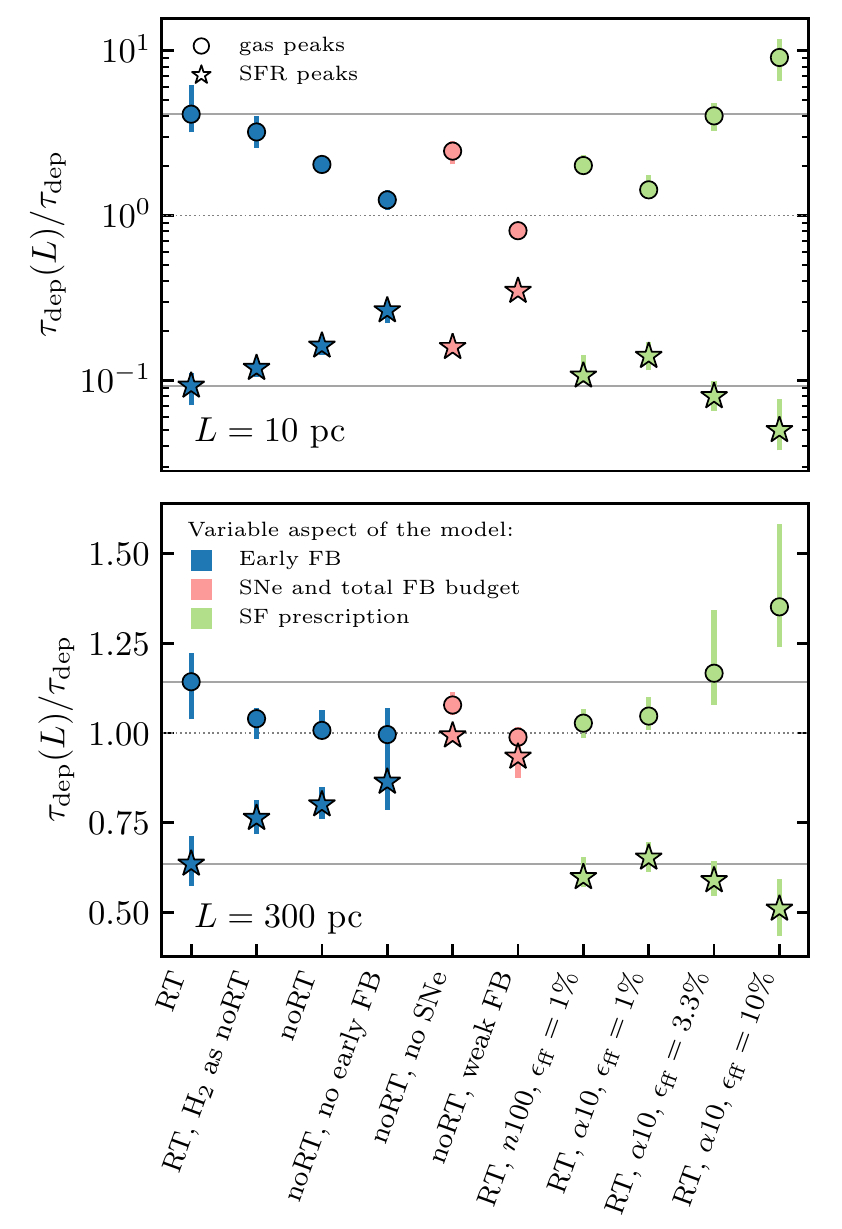}
\caption{\label{fig:tf-summary} The summary of the tuning fork sensitivity to variations of the star formation and feedback models shown in Figures~\ref{fig:tf-varfb} and \ref{fig:tf-varsf}. Different panels show the opening of the tuning fork sampled at scales $L=10\pc$ (top panel) and $L=300\pc$ (bottom panel), with circles and stars showing the branches that correspond to gas and SFR peaks, respectively. The error bars indicate snapshot-to-snapshot variation (16$^{\rm th}$--84$^{\rm th}$ interpercentile range). The points are colored according to the aspect of the model that was varied. Solid horizontal lines show the result of the fiducial RT simulation, while the dotted lines indicate the galaxy-averaged $\tauH2$. Note that the scale of the $y$-axis in the bottom panel is linear.}
\end{figure}

In the previous section, we presented a systematic exploration of the variation of the molecular depletion time on the choice of patch centers and scale---the tuning fork diagram---and its dependence on various aspects of galaxy modeling. Our findings are summarized in Figure~\ref{fig:tf-summary}, which shows how the opening of the tuning fork diagram changes in our runs with different star formation and feedback models. The two panels show the opening at small ($10 \pc$) and large ($300 \pc$) scales.

The results show that the diagram is indeed a sensitive probe of the feedback modeling in simulations, as was argued by \citet{Fujimoto.etal.2019}. Indeed, \citet{Jeffreson.etal.2020} demonstrated that their simulations, in which feedback is likely more efficient than in the simulations of \citet{Fujimoto.etal.2019}, produce a wider opening of the tuning fork diagram. The latter study, however, did not present a detailed comparison with observations or investigation of the factors that shape the form and opening of the diagram in their simulations. 

We confirm that the opening of the tuning fork diagram is quite sensitive to the strength of stellar feedback assumed in the simulations and is sensitive to the inclusion and duration of the ``early feedback'' stage. At the same time, we show that the inclusion of self-consistent modeling of RT and of H$_2$ abundance in simulations also contributes significantly to the widening of the tuning fork opening on $<100\pc$ scales.

Moreover, we  show that the tuning fork diagram is sensitive to the assumptions about the star formation efficiency per freefall time, $\epsff$, in star-forming regions. For example, Figure~\ref{fig:tf-varsf} shows that the simulations with different assumptions about $\epsff$ result in significantly different tuning fork diagram openings, even though they all reproduce the global SFR measured in NGC\,300. 

All of the above processes are part of the overall star formation--feedback cycle, so, to a certain degree, it is not surprising that they all affect depletion times in local patches of the ISM. Indeed, as we showed in our previous papers, the total gas and molecular depletion times explicitly depend on the feedback strength and $\epsff$ \citep[][]{Semenov.etal.2017,Semenov.etal.2018,Semenov.etal.2019}. 

For example, in the framework presented in these papers, the depletion time in a given patch of the ISM explicitly depends on the strength of stellar feedback (i.e., the amount of energy and momentum injected per unit of stellar mass formed), when the efficiency of star formation is sufficiently large to allow for efficient feedback (e.g., $\epsff\gtrsim 1\%$ for \Lstar~galaxies). The depletion time in such a regime is proportional to the strength of feedback quantified by the ``mass-loading factor'' $\xi$---the proportionality constant between the rate of star-forming cloud dispersal and its local SFR \citep[][]{Semenov.etal.2017,Semenov.etal.2018}. This dependence can explain the dependence of global depletion time on feedback strength demonstrated in a number of simulation studies \citep[see also][]{Hopkins.etal.2017,Orr.etal.2017,Semenov.etal.2018} and the  increase of $\tauH2$ (decrease of SFR for a given gas mass) with increasing feedback strength shown in Figure~\ref{fig:sfh-varfb}. However, this by itself does not explain the differential effect of the feedback strength on $\tauH2$ in gas-peak- and SFR-peak-centered patches and its increase with decreasing scale (i.e., the opening of the tuning fork diagram). The latter is likely due to the increasing scatter of $\tauH2$ with decreasing patch size within which it is measured. 

As shown by \citet{Feldmann.etal.2011}, such stochasticity is at least partly due to the fact that molecular gas measurements are \emph{instantaneous}, while estimates of SFR are necessarily averaged over a certain timescale. This allows the H$_2$ abundance to decrease locally due to effects of stellar feedback during the time period within which local SFR is averaged. Consequently, the apparent $\tauH2$ decreases as regions evolve from pre-star formation and early star formation stages to the late stage, when stars have already largely dispersed gas in their natal clouds  \citep[][]{Feldmann.Gnedin.2011,Kruijssen.Longmore.2014,Hu.etal.2016,Lee.etal.2016,Kruijssen.etal.2018}. It is reasonable to assume that the gas-peak-centered patches largely reflect pre- and early star formation stages of dense molecular gas, while patches centered on \Halpha~peaks correspond to late stages of star formation. This then manifests in the opening of the tuning fork diagram with decreasing scale.  In the context of the framework of \citet{Semenov.etal.2017}, the increasing stochasticity of $\tauH2$ with decreasing scale is due to averaging over different populations of ISM parcels in systematically different stages of their evolution in patches with different properties.  

Likewise, the origin of the differences in the molecular gas and SFR maps at different $\epsff$ that can be seen in Figure~\ref{fig:peakmaps-varsf} can be understood using the model of rapid gas cycling between star-forming and non-star-forming states in the evolution of an ISM parcel \citep{Semenov.etal.2017}. When feedback is efficient and quickly disperses star-forming regions, the lifetimes of such regions scale inversely with $\epsff$. Indeed, the lifetime of a given star-forming region is set by the total fraction of gas that needs to be converted into stars so that these stars can disperse the rest of the region, $\epsint \sim 1/\xi$. 
At higher $\epsff$, a given star-forming region reaches $\epsint$ sooner and therefore the lifetime of the region decreases. The \emph{instantaneous} fraction of star-forming gas in regions of any scale thus decreases, and the number of molecular gas peaks above a given sensitivity threshold becomes smaller \citep[see also the bottom panel of Figure~2 in][]{Semenov.etal.2018}.
On the other hand, the number of SFR peaks depends only on the formation rate of star-forming regions and the value of $\epsint$, which do not strongly depend on $\epsff$. The effect of varying $\epsff$  on the SFR peaks distribution is thus small. The difference in the population of gas and SFR peaks is responsible for the decorrelation of gas and SF peaks manifested in the opening of the tuning fork diagram with decreasing scale. The difference in their dependence on $\epsff$ discussed above thus provides a qualitative explanation for the trends of the tuning fork diagram with varying $\epsff$. 

Interestingly, we find that the star formation model with $\epsff$ varying with the local virial parameter of gas, as suggested by numerical simulations of star formation in molecular clouds \citep[e.g.,][]{Padoan.etal.2012,Federrath.2015,Kim.etal.2020.gmcsims}, provides the best match to the observed tuning fork in NGC\,300 among different models. This provides an additional motivation for such models in addition to strong theoretical motivation from molecular cloud simulations. 

Next, the dependence of $\tauH2$ on the averaging timescale in the SFR estimate also indicates that this timescale should be carefully considered and matched when comparing model results and observations of the tuning fork diagram. We also show that the opening of the tuning fork diagram depends on the specific choices in modeling sensitivity limits to the molecular gas detection (see Appendix~\ref{app:map-cuts} and Figure~\ref{fig:tf-var-H2-cuts} specifically). Thus, to make consistent comparisons, this sensitivity should also be modeled carefully. 

Finally, the tuning fork can generally be expected to depend on the average profile and characteristic size of the gas and SFR peaks on small scales, $L\lesssim 100$ pc, and carry information about peak clustering and large-scale structures, like spiral arms. The average profiles on small scales can generally depend not only on the selection criteria and sensitivity of observations but also on physical processes affecting the distribution of gas around the peaks (e.g., ISM turbulence, feedback, etc.). For example, the distribution of \Halpha~around the peaks can depend on specifics of dispersal of molecular gas in star-forming regions and anisotropy of the \Halpha~escape from these regions. Increased leakage of \Halpha~from star-forming regions will reduce the contrast of SFR peaks, thereby affecting the lower branch of the diagram, and will increase the cross correlation of gas peaks with \Halpha, thereby affecting the upper branch and reducing the opening of the tuning fork.
 
Overall, our results indicate that although a comparison of the model and observed tuning fork diagram indeed provides a sensitive test of galaxy formation models, the result depends on many different aspects of the model, not just feedback or the timescale for dispersal of star-forming regions. This means that a failure to match observations may not necessarily be due to any individual part of the model. At the same time, successful match can possibly come from a different combination of the modeled processes and thus may not uniquely identify the correct implementation of star formation, feedback, and ISM processes. This implies that a certain degree of degeneracy may exist, and thus comparisons with complementary observational statistics may be useful.

\subsection{Tuning Fork Diagram and the Distribution of Star-Forming Gas Lifetimes}
\label{sec:discussion:tsf}

\begin{figure}
\includegraphics[width=\columnwidth]{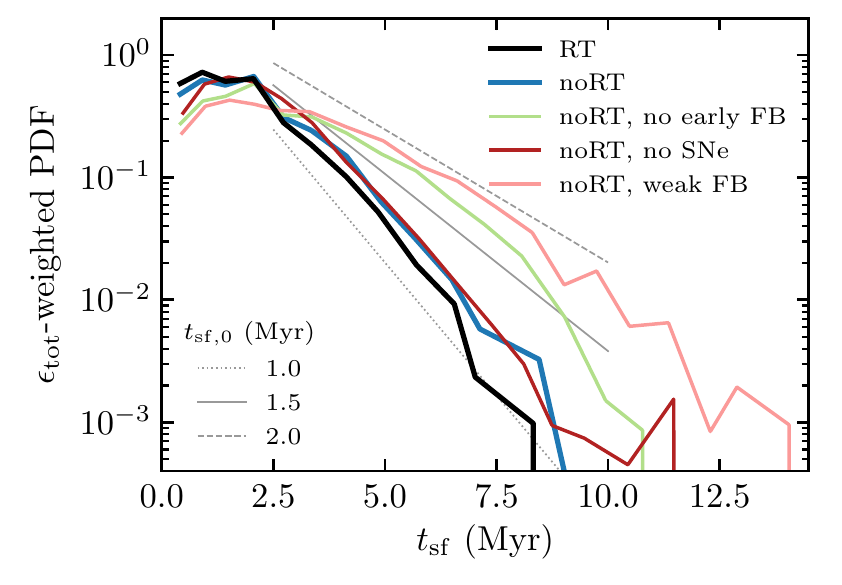}
\caption{\label{fig:tsf-varfb} Distribution of the residence times of gas tracer particles in the actively star-forming state with $\epsff > 1\%$. The contribution of each tracer trajectory is weighted by $\epsint = \int \epsff dt/\tff$, which corresponds to the total star formation efficiency over the trajectory. The high-$\tsf$ tail of the distribution is exponential, $\exp(-\tsf/t_{\rm sf,0})$, as shown by the thin line with the value of $t_{\rm sf,0}$ provided in the legend. This characteristic timescale and the short-$\tsf$ part of the distribution change only mildly, while the long-$\tsf$ tail depends strongly on the feedback model. The higher relative abundance of long-$\tsf$ gas translates to the narrower opening of the tuning fork on $<100\pc$ scales (see Figure~\ref{fig:tf-varfb}). }
\end{figure}

The lifetimes of star-forming regions estimated from observational measurements of the tuning fork diagram in galaxies are typically a few Myr \citep[][]{Chevance.etal.2020}, which is also the case for NGC\,300 \citep{Kruijssen.etal.2019}. This was interpreted as an indication that early feedback processes dominate in dispersing gas in star-forming regions and stopping star formation locally. Results of numerical experiments 
presented in Section~\ref{sec:results:fb} show  that although early feedback indeed dominates at averaging scales $<100$ pc, SN feedback does affect the opening of the tuning fork diagram at scales $\gtrsim 100$ pc. 

To clarify the relation with the lifetime of star-forming regions, we explored the distribution of star-forming gas lifetimes in runs with variations of feedback models discussed above using gas tracer particles and the analysis developed in \citet{Semenov.etal.2017,Semenov.etal.2018,Semenov.etal.2019}. Specifically, we populate our simulations with tracer particles that passively follow gas density in a Lagrangian manner \citep[][]{Genel.etal.2012} and track their evolution for $\sim 300 \Myr$. For each passage of a tracer particle through the star-forming state,\footnote{Although $\epsff$ varies continuously in our simulations, its exponential dependence on $\avir$ (Equation~\ref{eq:epsff-P12}) can be viewed as an effective $\avir$ threshold because $\avir$ in the ISM varies by orders of magnitude. For our analysis of gas tracers, we define the star-forming state as that with $\avir<10$. This threshold value of $\avir$ corresponds to $\epsff\sim1\%$, which accounts for $\approx 70\%$ of the total SFR in our fiducial simulation (see Figure~\ref{fig:epsff}).} we record the total contiguous time that the tracer spends in this state, $\tsf$, and the integral $\epsint = \int \epsff dt/\tff$ over the passage, which corresponds to the total star formation efficiency during $\tsf$. During the evolution, each tracer particle passes through multiple star-forming stages, and we record $\tsf$ and $\epsint$ for each such passage separately. Figure~\ref{fig:tsf-varfb} shows the probability density function (PDF) of $\tsf$ weighted by $\epsint$ in our simulations with different feedback models. 

A clear trend is evident from the figure: in simulations without early feedback or with the reduced momentum injection rate, the $\tsf$ distribution has a significant tail of high-$\tsf$ values. This is also reflected in the tuning fork diagram opening: simulations with more long-lived star-forming regions produce stronger correlation between dense gas and young stars, leading to a narrower tuning fork opening (see Figure~\ref{fig:tf-varfb}). Such regions are typically more massive and form stars more efficiently, which is consistent with our findings in Section~\ref{sec:results:sf} that the tuning fork is more sensitive to star-forming regions with the higher $\epsff$.

It is particularly interesting that the $\tsf$ distribution in the simulation without SNe is very close to that in our fiducial and ``noRT’’ simulations, while the global SFR in this run is a factor of $\sim$4 higher (see Figure~\ref{fig:sfh-varfb}). The similarity of $\tsf$ distributions indicates that early mechanical feedback can efficiently stop local star formation in all three runs, while the difference in the SFR implies that the \emph{instantaneous} fraction of star-forming gas is significantly smaller in the runs with SNe. The effects of SN feedback can be twofold: first, SNe can efficiently disperse dense gaseous regions on $>100\pc$ scales (compare these three runs in Figure~\ref{fig:maps-all}), and, second, the large-scale ISM turbulence driven by SNe can stabilize the disk and hinder the formation of such regions. In the context of our gas cycling model, both of these effects result in an increase of the gas residence time in the non-star-forming state, which, for a given $\tsf$, results in a smaller star-forming mass fraction and smaller SFR. As our results demonstrate, these effects also lead to a weaker correlation of dense gas and young stars (wider tuning fork) on the scales of hundreds of parsecs.

Another important conclusion from Figure~\ref{fig:tsf-varfb} is that short $\tsf$ dominate the distribution for all explored feedback models. The overall shape of the distribution is approximately exponential, with the characteristic timescale changing only weakly, between $\sim 1.5$ and $2\Myr$, as shown by the thin gray lines. Such values are shorter than the delay time between the onset of local star formation and first SNe ($\approx 3$ Myr) and are close to observational estimates of star-forming region lifetimes from age spreads in young star clusters \citep[e.g.,][see, for example, Figure~3 in the latter paper]{Reggiani.etal.2011,Kos.etal.2019} or from the tuning fork diagram itself \citep{Kruijssen.etal.2019,Chevance.etal.2020}. 

Our results show that it is indeed large relative abundance of the short-$\tsf$ regions relative to regions with long $\tsf$ in the simulations with efficient feedback that is responsible for their good match of the observed tuning fork diagram. At the same time, Figure~\ref{fig:tsf-varfb} shows that a distribution of lifetimes is expected and the observed tuning fork diagram is related to the characteristic timescale of this distribution \citep[see also][]{Jeffreson.etal.2021}.

\section{Summary}
\label{sec:summary}

We explored the sensitivity of the spatial correlation between dense gas and young stars to different aspects of star formation and feedback modeling in a suite of isolated sub-\Lstar~galaxy simulations that include an explicit treatment of RT and molecular chemistry. To quantify this correlation, we use the scale dependence of molecular gas depletion time in apertures centered on gas or SFR peaks---\emph{the tuning fork diagram} \citep{Schruba.etal.2010,Kruijssen.Longmore.2014,Kruijssen.etal.2018,Kruijssen.etal.2019,Chevance.etal.2020}. The bulk structural properties of our simulated galaxy are set to closely match those of NGC\,300 (see Section~\ref{sec:sims:ics} and Figures~\ref{fig:prof-vcZ} and \ref{fig:prof-H2SFR}), enabling a direct comparison with the recent observational measurements of the tuning fork diagram in that galaxy \citep{Kruijssen.etal.2019}.

In our simulation suite, we explored the effects of self-consistent modeling of the UV field and its effect on molecular gas, effects of early mechanical feedback and type II SNe, as well as different assumptions about local star formation efficiency, $\epsff$, in models both with variable $\epsff$ without any threshold for star-forming gas (motivated by simulations of turbulent star-forming regions) and with constant $\epsff$ and star-forming gas defined with a threshold in local virial parameter or density. The full list of the explored models is provided in Table~\ref{tab:sims} and visually summarized in Figure~\ref{fig:maps-all}. 

Our main results and conclusions can be summarized as follows:
\begin{enumerate}

\item The fiducial RT simulation reproduces the observed opening of the tuning fork in NGC\,300, indicating that the adopted star formation and feedback model is reasonably realistic (Figure~\ref{fig:tf-rt}). To our knowledge, this is the first time that this statistic was quantitatively reproduced in a galaxy formation simulation.

\item The success of the model is not due to any specific aspect of feedback; photodissociation of H$_2$, gas heating by the nonuniform UV field, and early mechanical feedback all contribute significantly to the tuning fork opening at $<100\pc$ scales (see Figure~\ref{fig:tf-varfb}). All these processes contribute to the ``early feedback'' phase and dominate over SN feedback on these scales. 

\item Nevertheless, we find that SN feedback does have a significant effect in shaping the tuning fork diagram on $\gtrsim 100\pc$ scales, especially its lower branch (see the bottom panel of Figure~\ref{fig:tf-varfb}).

\item We also find that the tuning fork diagram is quite sensitive to the value of star formation efficiency per freefall time, $\epsff$, with its opening increasing for larger $\epsff$ values (Figure~\ref{fig:tf-varsf}). This sensitivity is analogous to the effect of $\epsff$ on the tuning fork in an \Lstar~galaxy \citep[see Figure 11 in][]{Semenov.etal.2018}, and it can be used as a complimentary constraint on $\epsff$ in the regime where global SFR and depletion times are insensitive to the $\epsff$ value as is the case for NGC\,300 (see Figure~\ref{fig:sfh-varsf}). 

\item By comparing results of the runs with locally variable $\epsff$ and with a fixed $\epsff$ value, we find that the tuning fork diagram is sensitive to star-forming regions with the largest $\epsff$. Indeed, the tuning fork diagram from the simulation with variable $\epsff$ can be reproduced using a constant $\epsff = 3.3\%$ value, even though only $\approx 20\%$ of star-forming regions in the variable $\epsff$ run have $\epsff \geq 3.3\%$ (see Figure~\ref{fig:epsff}).

\item We explicitly show that the degree of opening of the tuning fork diagram  on $<100\pc$ scales is
directly related to the distribution of lifetimes of star-forming regions. The opening decreases when abundance of long-lived regions to short-lived regions is larger and vice versa (compare Figures~\ref{fig:tf-varfb} and \ref{fig:tsf-varfb}). 

\item The overall distribution of the star-forming gas lifetimes has a peak at $\tsf\approx 1$ Myr and can be approximated by an exponential PDF with the characteristic timescale of $\sim 1\text{--}2$ Myr at $\tsf>2$ Myr. 
These short typical values and wide distribution of $\tsf$ are qualitatively consistent with the timescales measured in simulations of an \Lstar~galaxy \citep{Semenov.etal.2017,Semenov.etal.2019}, which showed that short $\tsf$ values play a key role in setting long gas depletion times in galaxies and making star formation globally inefficient. 

\end{enumerate}

Our results indicate that the observed wide opening of the tuning fork diagram results from a combined effect of different aspects of star formation and feedback processes. Effects of each individual aspect are modest, comparable to the snapshot-to-snapshot variation of the tuning fork shown in Figure~\ref{fig:tf-rt}. Apart from star formation and feedback, multiple other parts of the model can have a comparable or even stronger effect. For example, the selection effects and sensitivity of molecular gas observations and the timescales probed by SFR indicators strongly affect the tuning fork opening (see Appendix~\ref{app:map-cuts}), and, therefore, they should be carefully considered for a consistent comparison with observations.

Overall, these results imply that the tuning fork diagram provides a stringent test not only of the feedback strength alone but of all aspects of the star formation--feedback cycle and of the details of forward-modeling observational analyses. This motivates both explorations of other statistical probes of  star formation, feedback, and ISM properties on subkiloparsec scales and improving the fidelity of simulations and realism of modeling of observational effects and analysis details.
From the observational side, probing the ISM structure down to small scales in a large number of star-forming galaxies such as NGC\,300 will provide additional stringent constraints on theoretical models and advance our understanding of the star formation--feedback cycle in galaxies.

\acknowledgments
We would like to thank M\'{e}lanie Chevance, Diederik Kruijssen, and Sarah Jeffreson for useful discussion of the tuning fork analysis and observational sensitivity of NGC\,300 measurements. We also thank the anonymous referee for a detailed review that helped to improve the manuscript. Support for V.S. was provided by NASA through the NASA Hubble Fellowship grant HST-HF2-51445.001-A awarded by the Space Telescope Science Institute, which is operated by the Association of Universities for Research in Astronomy, Inc., for NASA, under contract NAS5-26555.
This work was also supported by the NSF grants AST-1714658 and AST-1911111 and NASA ATP grant 80NSSC20K0512 and co-authored by Fermi Research Alliance, LLC under Contract No. DE-AC02-07CH11359 with the U.S. Department of Energy, Office of Science, Office of High Energy Physics. 
The simulations presented in this paper have been carried out using the Midway cluster at the University of Chicago Research Computing Center, which we acknowledge for support. Analyses presented in this paper were greatly aided by the following free software packages: {\tt yt} \citep{yt}, {\tt NumPy} \citep{numpy_ndarray}, {\tt SciPy} \citep{scipy}, {\tt Matplotlib} \citep{matplotlib}, and \href{https://github.com/}{GitHub}. We have also used the Astrophysics Data Service (\href{http://adsabs.harvard.edu/abstract_service.html}{ADS}) and \href{https://arxiv.org}{arXiv} preprint repository extensively during this project and writing of the paper.

\appendix

\section{Effect of feedback models on molecular gas and SFR profiles}
\label{app:prof-varfb}

\begin{figure}
\includegraphics[width=\columnwidth]{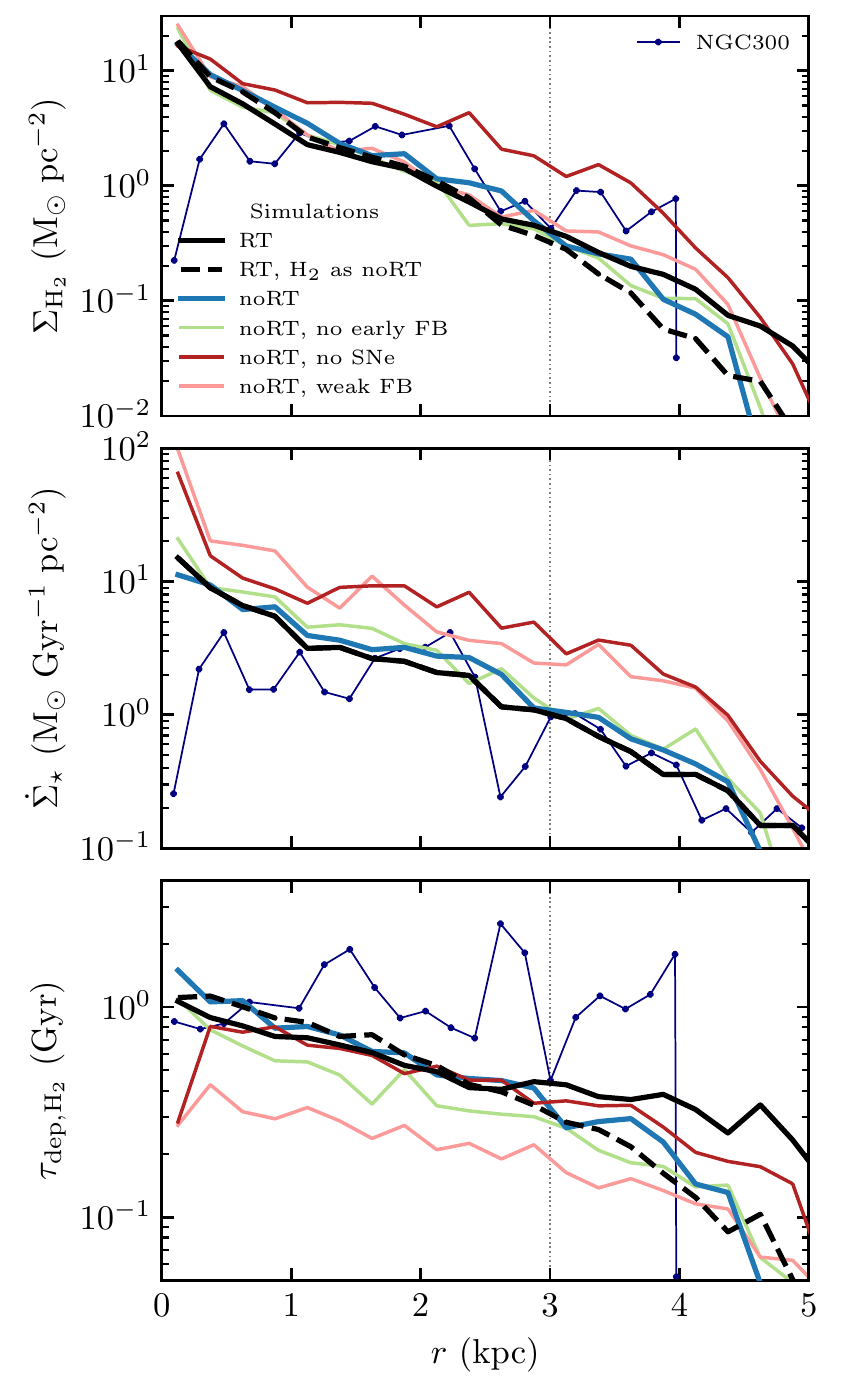}
\caption{\label{fig:prof-H2SFR-varfb} Variation of molecular gas (top panel), SFR surface density (middle panel), and molecular gas depletion time ($\tauH2 = \SH2/\SSFR$; bottom panel) profiles in simulations with different feedback models. The colors and line styles repeat those from Figures~\ref{fig:sfh-varfb} and \ref{fig:tf-varfb}. The simulated profiles are stacked over available snapshots, and the lines show running medians. For reference, the thin blue lines with markers show the observed profiles of NGC\,300 from \citet{Kruijssen.etal.2019}. The vertical dotted line marks the region where we apply our analysis, $R<3\kpc$.}
\end{figure}

Figure~\ref{fig:prof-H2SFR-varfb} demonstrates the sensitivity of $\SH2$, $\SSFR$, and $\tauH2 = \SH2/\SSFR$ profiles to variations of the feedback model explored in the paper. The molecular gas profiles show only a small variation inside the galactocentric radius of $R<3\kpc$, where we perform the analysis. The only exception is the run without SNe, where $\SH2$ is overestimated by a factor of 3--5 due to accumulation of dense gas on $>100\pc$ scales (see Figure~\ref{fig:maps-all}). The sensitivity of the $\SSFR$ profile is consistent with the trends in the global SFR shown in Figure~\ref{fig:sfh-varfb}: $\SSFR$ is weakly sensitive to variations of early feedback, but it increases by a factor of 3--5 in simulations with reduced total feedback budget and without SNe. Interestingly, the increase of $\SH2$ and $\SSFR$ in the simulation without SNe roughly cancels out, resulting in the same $\tauH2 = \SH2/\SSFR$ profile as in the fiducial simulation and, as a result, $\tauH2$ strongly deviates from the fiducial simulation only in the simulation with the reduced feedback budget.

\section{Effect of molecular gas and SFR selection}
\label{app:map-cuts}

\begin{figure}
\includegraphics[width=\columnwidth]{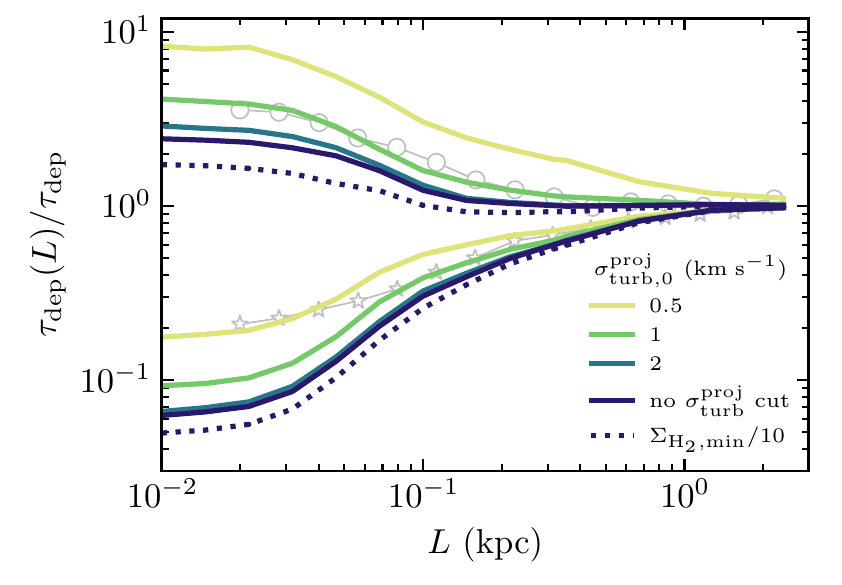}
\caption{\label{fig:tf-var-H2-cuts} Dependence of the tuning fork on the sensitivity cuts applied to the simulated $\SH2$ map (see Section~\ref{sec:sims:tf}). Line colors show the dependence on the value of the prefactor on the right-hand side of Equation~(\ref{eq:SH2-sigmamax}), with a darker color corresponding to a less stringent cut that includes more H$_2$ mass. The darkest blue line shows the case when the velocity cut is not applied at all. In addition, the dotted line shows the case without a $\sigma_{\rm turb}^{\rm proj}$ cut and assuming a 10 times better sensitivity to $\SH2$ than the fiducial value: $\SH2 > 1.3 \Msunpc2$. This last case includes almost all H$_2$ mass produced in the simulation. Simulation results are stacked over available snapshots, and lines show the median values (see Section~\ref{sec:sims:tf}). For reference, thin gray lines with markers show the observed relation in NGC\,300 from \citet{Kruijssen.etal.2019}. The differences in the tuning fork diagram for various choices of the sensitivity cuts are comparable in magnitude to the effects of the star formation and feedback model variations shown in Figures~\ref{fig:tf-varfb} and \ref{fig:tf-varsf}, and therefore, such sensitivity limits should be modeled carefully for consistent comparisons of model results to observations. }
\end{figure}

\begin{figure}
\includegraphics[width=\columnwidth]{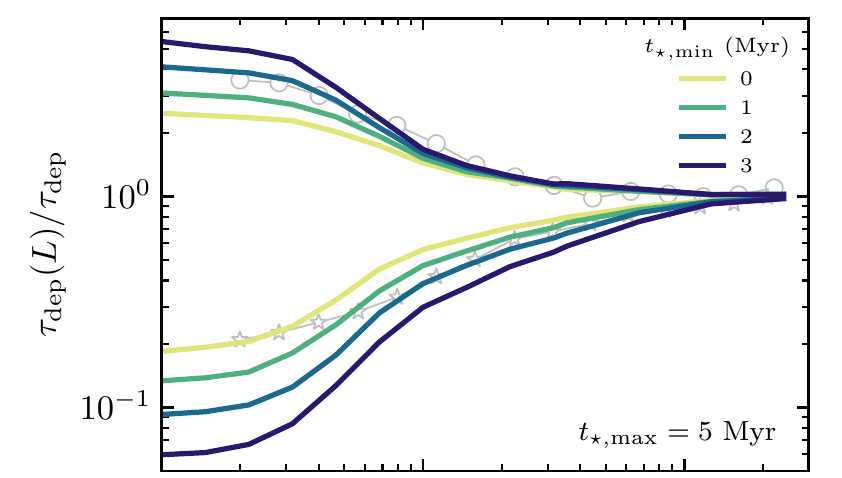}\\
\includegraphics[width=\columnwidth]{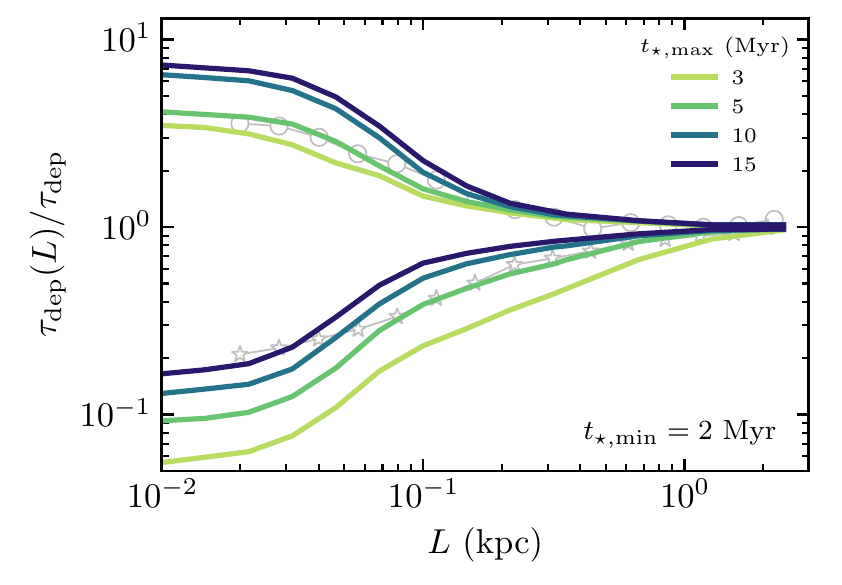}
\caption{\label{fig:tf-var-t} The effect of the age cuts of stellar particles used to construct the $\SSFR$ map (see Section~\ref{sec:sims:tf}). The top and bottom panels show the dependence on the lower and upper age cuts, respectively (i.e., the duration of the embedded stage and the timescale over which a stellar population can be seen in \Halpha), while keeping the other cut at the fiducial value shown at the lower right corner of each panel. Simulation results are stacked over available snapshots and lines show the median values (see Section~\ref{sec:sims:tf}). For reference, thin gray lines in both panels show the observed relation in NGC\,300 from \citet{Kruijssen.etal.2019}.}
\end{figure}

In this appendix, we illustrate the dependence of the tuning fork diagram on the adopted $\SH2$ sensitivity limits and the ages of star particles used to generate the $\SSFR$ maps. 

Figure~\ref{fig:tf-var-H2-cuts} shows the sensitivity of the tuning fork to the selection cuts applied to the simulated $\SH2$ map (Equations~(\ref{eq:SH2-min}) and (\ref{eq:SH2-sigmamax})). Explored variations are detailed in the figure caption.
As the selection cuts become less stringent and include more H$_2$ in the $\SH2$ map, both branches of the tuning fork shift downward and the tuning fork opening changes only mildly. 

Figure~\ref{fig:tf-var-t} shows the effect of the stellar particle ages used to select stellar populations visible in \Halpha. The top panel shows the dependence on the lower age cut that approximates the duration of the embedded stage of star formation, when \Halpha~emission is absorbed by the natal star-forming region. Observational estimates typically suggest that this phase can last for a few Myr. As expected, the tuning fork opening widens as this timescale increases because older stars are expected to be less correlated with dense gas. Interestingly, however, the opening of the tuning fork remains rather wide even when we set this timescale to 0 and use all young stars in our analysis.

The bottom panel shows the effect of the upper age cut on the tuning fork. This cut corresponds to the typical timescale over which \ion{H}{2} regions around O and B stars from a single-age population are expected to emit \Halpha. As this age cut is increased, the $\SSFR$ map includes more and more older stars that correlate with dense gas more weakly. As a result, the effect on the tuning fork is analogous to the effect of increasing $\sigma_{\rm turb}^{\rm proj}$ cut in the $\SH2$ map in Figure~\ref{fig:tf-var-H2-cuts} that leads to a larger fraction of diffuse and turbulent molecular gas that correlates with young stars more weakly. The direction of the effect is opposite to that shown in Figure~\ref{fig:tf-var-H2-cuts} because $\SSFR$ enters in the denominator of $\tauH2$.

The results shown in Figures \ref{fig:tf-var-H2-cuts} and \ref{fig:tf-var-t} demonstrate that the differences resulting from varying the assumptions about the sensitivity limits or the duration of embedded and \Halpha-bright stages of star formation are comparable to the effects of different assumptions about star formation and feedback processes. This implies that the sensitivity cuts and star formation timescales should be considered carefully when model results are compared with observational measurements of the tuning fork diagram.

\bibliographystyle{aasjournal}
\bibliography{}

\end{document}